\documentclass[fleqn,usenatbib]{mnras}

\usepackage{newtxtext,newtxmath}
\usepackage[colorinlistoftodos]{todonotes}

\usepackage[T1]{fontenc}

\DeclareRobustCommand{\VAN}[3]{#2}
\let\VANthebibliography\thebibliography
\def\thebibliography{\DeclareRobustCommand{\VAN}[3]{##3}\VANthebibliography}

\usepackage{subfig}

\usepackage{graphicx}	
\usepackage{amsmath}	
\usepackage{newtxtext,newtxmath}	
\usepackage{multirow}
\usepackage{tabularx}
\usepackage{arydshln}
\usepackage{multicol}

\title[Trajectory uncertainty of Comet Interceptor targets]{Uncertainty estimation of incoming comet trajectories with a thermophysical non-gravitational force model for Comet Interceptor encounter planning}

\author[N.~Attree et al.]{
N.~Attree$^{1,}$\thanks{E-mail:attree@iaa.csic.es},
P.~de Boer$^{2,}$$^{1}$,
P.~Guti{\'e}rrez$^{1}$,
P.~Lacerda$^{3}$,
M. K{\"u}ppers$^{4}$,
D. Dirkx$^{2}$
\\
$^{1}$Instituto de Astrofísica de Andalucía (CSIC), Glorieta de la
Astronomía s/n. 18008 Granada, Spain\\
$^{2}$Luchtvaart- en Ruimtevaarttechniek, Technische Universiteit Delft, Kluyverweg 1, 2629 HS Delft, The Netherlands\\
$^{3}$Leiden Observatory, Leiden University, PO Box 9513, NL-2300 RA Leiden, The Netherlands\\
$^{4}$European Space Agency (ESA), ESAC, Camino Bajo del Castillo s/n, 28692 Villanueva de la Cañada, Madrid, Spain\\
}

\date{Accepted XXX. Received YYY; in original form ZZZ}

\pubyear{2024}

\begin{document}
\label{firstpage}
\pagerange{\pageref{firstpage}--\pageref{lastpage}}
\maketitle

\begin{abstract}

The European Space Agency Comet Interceptor mission is designed to fly-by an as-yet undetermined incoming long-period or dynamically-new comet, which will require an accurate prediction of its trajectory. Cometary orbits are perturbed from purely gravitational trajectories by non-gravitational accelerations (NGAs) stemming from their outgassing, however, and these must be taken into account in the navigation. We here test the ability of ground-based astrometry and standard NGA-fitting techniques to predict the position and velocity of incoming comets perturbed by a Monte Carlo sampled thermophysical acceleration model. We find that, due to the difficulties of predicting NGA evolution as well as the limitations of reasonably simulated astrometry, predicted accuracies are unlikely to meet the mission goals of 1000 km and 1 m s$^{-1}$ in position and velocity, sixty days before intercept. We instead recommend expanded uncertainties of $10,000$ km and several m s$^{-1}$. Our NGA fits struggle to reproduce the reference accelerations and trajectories when outgassing is rising steeply at a few AU-incoming, but with more observations up until perihelion, do eventually converge, leading to good estimates of many of the underlying acceleration parameters even for half an orbital arc.

\end{abstract}



\section{Introduction}
\label{sec:introduction}
The European Space Agency's Comet Interceptor (CI) Mission is set to fly by and conduct in situ measurements on a yet-to-be-determined dynamically new, long-period comet (DN LPC) between 2029 and 2034 \citep{jones-2024}. DN LPCs are among the most pristine remnants of the solar system formation, originating predominantly from the Oort Cloud and entering the inner solar system on hyperbolic or near-parabolic trajectories \citep{Rickman2010}. To conduct a successful encounter of the spacecraft with such a target, two requirements are stated on the uncertainty of the predicted cometary state. Sixty days prior to intercept, the position uncertainty must meet $\sigma_{pos}\leq1000~km$ with a velocity uncertainty of $\sigma_{vel}\leq1~ms^{-1}$. Here, we asses the achievability of these uncertainty requirements given typical orbit-determination techniques with noisy astrometric measurements and non-gravitationally perturbed comet orbits.

When DN LPC approach the inner solar system,  volatiles 
start to sublimate, creating Non-Gravitational Accelerations (NGA), perturbing the trajectory. A widely used approach to model these effects was introduced by \cite{marsden-1973}, which represents the NGA as an orbital-averaged perturbation estimated along with the state of the comet. In this formulation, the NGA are expressed as:
\begin{equation}\label{Marsden_symmetric}
    \centering
    \vec{\mathbf{a}}_{NG}=g(r)A_i~,\quad A_i=const~[AU/day^2],\quad i=1,2,3,
\end{equation}
where $A_i$ are constants estimated for each comet for the accelerations in the cometocentric RTN (radial, tangential, normal) frame; $A_1$, $A_2$, and $A_3$ correspond to the components in the radial direction from the comet to the Sun, the direction tangential to its orbit, and the direction along the orbit normal, respectively. The heliocentric distance dependence of the outgassing accelerations is described via the dimensionless semi-empirical function g(r):
\begin{equation}\label{g(r)powerlaw}    
    g(r)=\alpha(r/r_0)^{-m}[1+(r/r_0)^n]^{-k},
\end{equation}
which scales the $A_i$ parameters according to a water-production-driven, symmetric outgassing law about perihelion. The standard water production scaling parameters are $m = 2.15$, $n = 5.093$,  $k = 4.6142$, $r_0 = 2.808$ AU, and $\alpha = 0.1112620426$, following the \textit{Jet Propulsion Laboratory (JPL)} convention that normalises the power law such that $g(r=1 AU) = 1$. To account for the asymmetric outgassing behaviour relative to perihelion displayed by some comets, \cite{yeomans-1989} introduced a time-shift parameter $\tau$, modifying the sublimation law as $g(r)\rightarrow g(r') \quad\text{with}\quad{r'=r(t')=r(t-\tau)}$. In this context, $\tau<0$ corresponds to pre-perihelion while $\tau>0$ corresponds to post-perihelion peak activity.

\citet{marsden-1973} and \citet{yeomans-1989}-style models are typically fitted to periodic comets with multiple apparitions, with, as mentioned above, the $A_i$ terms representing orbitally-averaged accelerations. Instantaneous acceleration vectors of real, outgassing comets are therefore not necessarily represented by these NGA curves (see, for example, \citealp{Farnocchia2016}). Hence, for precise orbit determination (POD) of particular comets of interest, more realistic models are often used, such as the rotating-jet model \citep{Chesley2005}. Here, one or more `jets' are placed at specific locations on an assumed spherical nucleus so that their directional vector can be used to calculate more realistic NGA curves. This requires knowledge of, or places constraints upon, the pole-axis orientation of a particular comet, but the total `power' of the jets is still scaled by a heliocentric dependence such as Eqn.~\ref{g(r)powerlaw}. Even more realistic NGA models compute accelerations directly from thermophysical outgassing, typically over an assumed triaxial ellipsoid-shaped nucleus \citep{maquet-2012} or a full comet shape-model (see, e.g.~\citealp{attree-2024}). These can produce more accurate ephemerides compared to \cite{marsden-1973} models, but require much more information about the nucleus physical properties and/or trajectory than is typically available from ground-based observations (e.g.~for the latter, detailed measurements of comet 67P/Churyumov-Gerasimenko's nucleus and trajectory made by ESA's Rosetta mission; \citealp{LasagniManghi2024}).

Whilst thermophysical models improve physical realism, they are limited to comets with known physical properties (sizes, shapes, spins, activity patterns, etc.~), so that typical precise orbit determination techniques for LPCs still employ the Marsden-formulated NGA. Orbit fits, including the estimation of the Marsden accelerations provided by JPL, are typically generated via pre- and post- perihelion arc observations. This information is not available for the CI mission, as the flyby will likely be performed pre-perihelion, or very shortly after it  \citep{jones-2024}. These limitations are apparent in \textit{The catalogue of cometary orbits and their dynamical evolution (CODE)} from \cite{krolikowska-2020}, which provides orbit fits for nearly 300 LPCs based on different observation arcs. A consistent observation is that orbit fits relying on pre-perihelion observations yield significantly different estimated Marsden accelerations compared to those with an extended data arc. This suggests that pre-perihelion data only provide weak constraints on the estimation of the Marsden accelerations. 

\begin{figure}
    \centering
    \includegraphics[width=1\linewidth]{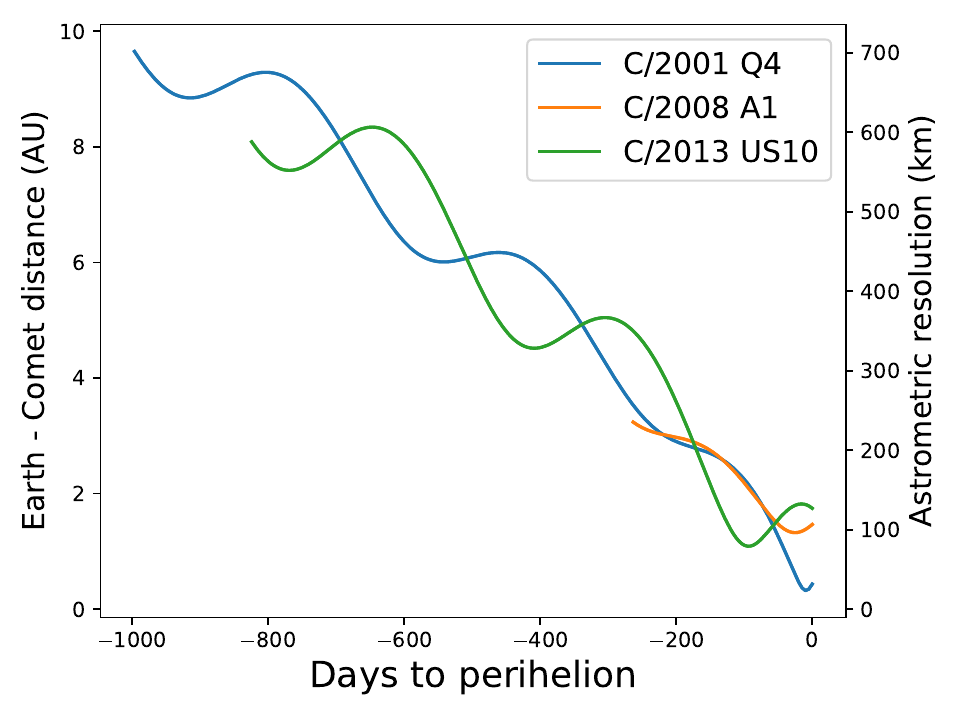}
    \caption{Earth - comet distance for the three virtual target comets in AU. The right-hand scale shows the equivalent linear resolution at the comet, assuming 0.1 arcseconds astrometry (see Sect.~\ref{subsubsec:Obsmodel}).} 
    \label{fig:rE}
\end{figure}

In this paper, we aim to assess the effect of the Marsden and thermophysical Non-Gravitational Force (NGF) models on the POD routines of DN LPC and evaluate the reachability of the mission requirements regarding the cometary state. This is done via the analysis of historical LPCs C/2001 Q4, C/2008 A1, and C/2013 US10, which have been selected as example targets for the CI mission by \cite{snodgrass-2025}. Osculating orbital elements for these comets are shown in Table \ref{tab:comets}, and their Earth - comet ranges are plotted in Fig.~\ref{fig:rE}. In Section \ref{sec:method} the methodology behind the orbit fitting routines is described, followed by the implementation of two different NGF models. Section \ref{sec:results} shows the results of the orbit fitting routines, with discussion in Section \ref{sec:discussion}, and conclusions drawn in Section \ref{sec:conclusions}. Although this study is motivated by the requirements of the Comet Interceptor mission, the proposed methodology is broadly applicable to the prediction of newly discovered long-period comet trajectories. It provides a framework for quantifying the interplay between observational uncertainty and non-gravitational forces, and may therefore support survey planning, follow-up observation strategies, target selection for future rendezvous missions, and studies of cometary orbit predictability more generally.

\begin{table}
\caption{Comet Interceptor virtual target osculating elements (truncated from JPL): semi-major axis, eccentricity, perihelion distance, inclination, longitude of ascending node, argument of perihelion, and time of perihelion passage. Note, all three comets are hyperbolic so have negative semi-major axes.}
\label{tab:comets}
\begin{tabular}{lccc}
\hline
& C/2001 Q4 & C/2008 A1 & C/2013 US10\\
\hline
a (AU) & -1398.092 & -4655.992 & -2937.555 \\
e & 1.00069 & 1.00023 & 1.00028 \\
q (AU) & 0.962 & 1.073 & 0.823 \\
i ($^{\circ}$) & 99.643 & 82.550 & 148.878 \\
$\Omega$ ($^{\circ}$) & 210.278 & 277.884 & 186.145  \\
$\omega$ ($^{\circ}$) & 1.204 & 348.473 & 340.359  \\
T & 2004-05-15 23:10 & 2008-09-29 02:59 & 2015-11-15 17:19 \\
\hline
\end{tabular}
\end{table}

\section{Method}
\label{sec:method}
In this section, the POD routines of cometary trajectories perturbed by NGF models are described. The two different NGF models are the Marsden model of Eqn.~\ref{Marsden_symmetric} (Model I), and the thermophysical NGF model (Model II) that is described in details in in the second subsection.

\subsection{Precise Orbit Determination (POD)}\label{subsec:POD}
The precise determination of cometary trajectories depends on the simulated environment, the force models acting on the comet, and the quality of astrometric observations used during the estimation process \citep{2004come.book..137Y}. In this subsection, these dependencies are explained. This is followed by the state estimators and how the trajectory uncertainty is measured. The POD routines are performed with the open-source Tudat(py)\footnote{https://docs.tudat.space/en/latest/} simulator \citep{Dirkx2022}.

\subsubsection{Environment \& Dynamics}
\label{dynamics}
The environment comprises the major planets of the Solar System. Kernels within Tudat\footnote{tudatpy.interface.spice.load\_standard\_kernels} were loaded to provide IAU- orientation and size/shape for natural bodies, IMMCE INPOP/NOE-derived masses and rotation models of solar system planets, and the NAIF \citep{acton-1996} leap-second kernel for TAI–UTC time conversions. Planetary ephemerides were obtained from the JPL DE441 SPICE kernel. In addition, custom SPICE kernels were generated for the three reference comets via the NASA JPL Horizons API. 

The dynamical model includes a point mass attraction for each considered body. For the Sun, Jupiter and Saturn, first-order relativistic effects with the Schwarzschild correction \citep{Pavlis2002Chapter10} have been applied. Additionally, the comet itself is under the effect of an NGF model, described by the Marsden parametrisation for Model I, or the thermophysical model found in Section \ref{subsec:Thermophysical_Model} for Model II. Note that none of the three comets examined here have a measured NGA time-offset, $\tau$, so that Model I uses only the symmetric NGA curve of  Eqn \eqref{Marsden_symmetric} \citep{marsden-1973}, and not the asymmetric curve of \citet{yeomans-1989}.

The orbits are integrated between the first observation and the perihelion time, as provided by the JPL SBDB, using a fixed-timestep Runge-Kutta 6 method. Note that this results in observing arcs of differing lengths for each comet (see Fig.~\ref{fig:rE}). The simulation is initialised in the heliocentric Ecliptic J2000 reference frame. The fixed timestep is $\Delta t=1$ day, as a compromise between numerical accuracy and computational speed. This is tested by comparing to an integration with $\Delta t/2$, with the maximum position error compared to the higher resolution model found to be  below 0.1 meters; at the boundary between floating-point and truncation numerical errors.


\subsubsection{Synthetic observations}\label{subsubsec:Obsmodel}
Synthetic observations are generated from reference (“truth”) trajectories propagated with the dynamical models described above, i.e.~by converting heliocentric truth-orbit positions in the J2000 Ecliptic frame, $x_{t},y_{t},z_{t}$, to right-ascensions and declinations as viewed from an observatory on Earth. For the Marsden formulation, a single truth trajectory is defined. In contrast, the thermophysical model permits multiple truth trajectories depending on the number of NGA-curve samples.

The observation survey and precision are modelled after the expected performance of the Vera C. Rubin Observatory’s Legacy Survey of Space and Time (LSST). Under an ideal survey, the LSST is expected to increase the discovery rate of long-period and hyperbolic comets by at least a factor of two \citep{LPC_LSST_DETECTION}. The design astrometric precision is 10 milli-arcseconds \citep{LSST_DESIGN}; to account for the signal-to-noise-ratio, an astrometric noise of 0.1 arcseconds is adopted as the noise level in the synthetic observation model. Initial LSST results with cometary targets suggest that this is reasonable, with \citealp{Chandler_2026} adopting a `conservative' 0.25 arcsecond astrometric noise for processing 3I/ATLAS detections, with the final post-fit position uncertainty deemed highly precise at 70 milli-arcseconds. We test the effects of increased astrometric noise in Sect.~\ref{Results:astrometry}, below.

A simplified scenario is assumed in which the observatory is located at the geocenter. This simplification removes topocentric effects, allowing us to determine the number of required observations of an observatory with a certain noise level. This approach also explains the exclusion of historical data from the IAU Minor Planet Center (MPC). Observations are generated once per Julian day from the initial detection defined by the JPL SBDB until an integer Julian day before perihelion. This results in a total number of observations of 996, 264 and 824 for C/2001 Q4, C/2008 A1 and C/2013 US10, respectively. These observations were given an astrometric noise of 0.1 arcseconds to right ascension and declination using a Gaussian distribution. 

\subsubsection{POD - initial state estimators}
Once the synthetic observations of each truth orbit have been created, the orbit estimation (POD) is made from them. The initial state of the determined orbit is estimated via an iterative batch least squares (LSQ) algorithm, with both parameter estimation and covariance propagation performed via the estimation framework in Tudat. An observation weighting matrix is included to account for the measurement errors which are assumed equivalent to the astrometric noise above. The weighting matrix is assumed to be uncorrelated and of constant value for  all the synthetic observations. The estimation is iteratively performed by removing subsets of synthetic observations with respect to the perihelion time until a certain minimum number of observations is reached. Three different dynamical models are considered for the estimated orbits:

\medskip

\textit{Case A} consists solely of $N$-body gravitational interactions. 
The estimated parameter vector contains only the initial Cartesian state; the state transition matrix, therefore, contains only the gravitational partial derivatives to the position and velocity. The goal of this estimator is to determine whether the low-noise astrometric observations are sensitive enough to reveal the effect of unmodelled NGAs during the estimation and trajectory uncertainty measurement.


\medskip

\textit{Case B} estimates, in addition to the initial Cartesian state, the Marsden non-gravitational $A_i$ parameters under the assumption of symmetric outgassing around perihelion. The variational equations are extended by including the partial derivate of eqn. \eqref{Marsden_symmetric} with respect to the $A_i$ parameters. As discussed in the introduction, $A_i$ estimation based on pre-perihelion observation provides weak constraints. The objective of Case B is to assess whether this is enough, combined with the low astrometric noise of LSST, to reliably estimate the state at perihelion.


\medskip

\textit{Case C}, the time-offeset parameter $\tau$ is additionally estimated, 
such that the sublimation function becomes $g(r')$, where $r'=r(t-\tau)$ as per \cite{yeomans-1989} description. The partial derivative is taken with respect to $\tau$ and the $A_i$ parameters. In the chain-rule an additional parameter $\overset{.}{\tau}$ appears, which is the change of the asymmetric outgassing in day/day. In this study, $\overset{.}{\tau}$ is set to be zero and thus not estimated. This is because $\overset{.}{\tau}$ is related to comets with multiple apparitions.

The cometary state $r'$, $\mathbf{r'~\texttt {and}~v'}$ is retrieved from a pre-propagated reference trajectory governed by the $N-$body dynamics. To test this method, the three comets have been given several artificial $\tau$ values up to ±100 days. We find that during the integration of the comet, using the dynamics in section \ref{dynamics} with the \citet{yeomans-1989} NGF model, the heliocentric powerlaw g(r) peaks a maximum of -0.0342, -0.0422 and 0.0549 days earlier or later than the given $\tau$ value for C/2001 Q4, C/2008 A1 and C/2013 US10, respectively. Additionally, this $\tau$ offset is independent of the given $\tau$ value. The Case C estimator's objective is to determine the $A_i$ and $\tau$ parameters simultaneously for the thermophysical perturbed trajectories. 

With two different truth orbit NGF models (I and II), and three different estimated orbit models (Cases A, B, and C), we perform a number of different integrations and PODs that are summarised in Table \ref{tab:runs}. 

\begin{table}
\caption{Summary of simulations performed, showing designations of the different NGF models used in the underlying “truth” orbits, and the orbits estimated from them by the POD procedure. Sym.~Marsden refers to the symmetric Marsden model of \citet{marsden-1973}, while asym.~Marsden refers to the asymmetric model of \citet{yeomans-1989}, and the thermophysical model is described in Sect.~\ref{subsec:Thermophysical_Model}. Note that Model I, Case C was not run as non -applicable.}
\label{tab:runs}
\begin{tabular}{c|ccc}
\hline
Truth orbit Model & \multicolumn{3}{c}{Estimated orbit Case}\\
& Pure gravity & sym.~Marsden & asym.~Marsden \\
\hline
sym.~Marsden & IA & IB & - \\
Thermophysical & IIA & IIB & IIC \\
\hline
\end{tabular}
\end{table}

\subsubsection{Trajectory uncertainty analysis}
The formal uncertainties of the estimated parameter vectors are captured in the covariance matrices. By using a multivariate normally distributed Monte Carlo sampling of a covariance matrix around the estimated parameter vector, we can generate a population of initial state vectors and Marsden NGA parameters, which are then propagated forward to the time of Comet Interceptor's encounter (note that this is a standard covariance propagation for linearly propagating uncertainties). For simplicity, we here set the encounter at the comets' perihelions, rather than at their node crossings as will be targeted in the real case \citep{jones-2024}, noting that two are within ten days of one other for the three virtual target comets studied here.

This family of possible solutions are then tested against the ``truth'' trajectories by comparing their geometric deviations at encounter (from here on; perihelion) to find the true trajectory error,
\begin{equation}
\Delta=\sqrt{(x_{i}-x_{t})^{2}+(y_{i}-y_{t})^{2}+(z_{i}-z_{t})^{2}},
\label{eq:Delta}
\end{equation}
for the heliocentric coordinates of each trajectory in the family, $i$, and the truth orbit $t$. A distribution of  $\Delta$ over many samples is then generated. For Model I, the distribution is over the family of 1000 Monte Carlo samples of the covariance matrix, with a single truth orbit. For Model II, $\Delta$ is distributed over the family of Monte Carlo samples of the covariance matrix for each of the separately Monte Carlo sampled NGA curves (e.g. 1000 sampled NGA curves each having 500 Monte Carlo samples). In the results section below, we plot the median error, $\tilde{\Delta}$, with error bars denoting the interquartile range (i.e.~$\tilde{\Delta}\pm25\%$ of the distribution), as robust measures of the average true uncertainty and its spread.

\subsection{Thermophysical model}
\label{subsec:Thermophysical_Model}

\subsubsection{Physical model}
\label{subsubsec:Physical_Model}

A physical model of the comet nucleus is needed to apply the thermal model to. Typical thermophysical NGF models (e.g.~see \citealp{davidsson-2004, Davidsson2005} and \citealp{maquet-2012}) define a nucleus size and shape, often in terms of a triaxial ellipsoid, with a set pole-axis orientation, and certain thermal parameters, sometimes including a thermal lag-time. They then typically sample a range of orientations, ellipsoid length-ratios, and surface activity patterns, within some range of reasonable parameters based on observations of a particular comet. Since we are concerned here with a theoretical study of as yet unknown comets, and merely want a range of plausible NGF curves, we choose to use a real comet shape, with a set pole-axis orientation, and calculate realistic outgassing forces from various surface activity patterns on this.

For this purpose we use the very well studied example of 67P, for which detailed shape models and rotational information are available. Following the method of \cite{attree-2019}, we use the {\it SHAP7} shape model \citep{Preusker17} with 125k facets, spin period $P=12.4$ hrs and pole-axis orientation $\rm{RA}=69.54^{\circ}, \rm{Dec} = 64.11^{\circ}$. In order to randomly sample different surface activity patterns, we divide the facets up into the regions defined in \citet{Thomas15}. Although these regions were identified based on differing morphological characteristics, we do not make any assumptions or inferences about their relative activity levels here; we merely use this surface splitting as a convenient way of defining 26 roughly similarly sized regions to sample in our Monte Carlo model. The physical comet model is placed onto the orbits of the three studied comets, using their SPICE kernels, with the above pole-axis orientation ensuring a random orientation relative to their orbital planes which differs for each of the three (obliquities of $83.1^{\circ}, 95.1^{\circ}$, and $124.6^{\circ}$ for C/2001 Q4, C/2008 A1, and C/2013 US10, respectively).

\subsubsection{Thermal model}
\label{subsubsec:Thermo_Model}
The thermophysical model used is based on \cite{attree-2019}; the energy balancing of each facet \textit{i} is given by Eqn. \eqref{eq:Thermal_balance}:
\begin{equation}\label{eq:Thermal_balance}
    (1-A_b)\frac{F_{\odot}}{r^2}\cos{\psi_i}=\epsilon\sigma T^4_{i}+L_{H_2O}Z_{H_2O}(T_i),
\end{equation}
where the left-hand side describes the energy intake due to the solar radiation with the Bond albedo $A_b$ = 0.0119 \citep{fornasier-2015}, $F_\odot$ the Solar constant $1370~Wm^{-2}$ at 1 AU, $r$ the Heliocentric distance and $\psi$ the zenithal angle. On the right-hand side of Eqn. \eqref{eq:Thermal_balance}, the energy loss is given by the thermal radiation of each facet into space using the assumed infrared emissivity $\epsilon=0.95$, $\sigma$ the Stefan-Boltzmann constant $Wm^{-2}K^{-4}$, $T_i$ the temperature of each facet. The second term is the latent heat loss due to water sublimation, where $L_{H_2O}=2.66\cdot10^6Jkg^{-1}$ is the latent heat of water ice sublimation at 200K. The mass sublimation rate $Z_{H_2O}$ $(kgs^{-1} m^{-2})$ is given in Eqn.~\eqref{eq:Water_sublimation}:
\begin{equation}\label{eq:Water_sublimation}
    Z_{H_2O}(T_i) = A~\text{exp}(\frac{-B}{T_i})\sqrt{{\frac{M_{H_2O}}{2\pi RT_i}}}~,
\end{equation}
with the two constants $A=3.561012$ and $B=6141.67$ from \cite{panale-1984}, the molecular mass $M_{H_2O}=0.018kg$ and finally the gas constant $R=8.3144598JK^{-1}mol^{-1}$. 
\subsubsection{Water production and NGA}
The resultant mass sublimation is converted to a water production rate per region via Eqn. \eqref{eq:Water_production},
\begin{equation}\label{eq:Water_production}
    Q_{H_2O}^{Region}=\sum_i~EAF~Z_{H_2O}(T_i)~S_i~,
\end{equation}
by summing over facets $i$ with areas $S_i$ forming a region on 67P, with EAF being the effective active fraction. The recoiling force of the mass sublimation by facet $i$ is calculated via Eqn. \eqref{eq:QtoF} \cite{attree-2024}, 
\begin{equation}\label{eq:QtoF}
    F_i = -\eta~EAF~Z_{H_2O}(i)\sqrt{\frac{8RT_{dust}}{\pi M_{H_2O}}},
\end{equation}
where $\eta$ is the momentum transfer coefficient \cite{crifo-1988}. From \cite{attree-2019}, $\eta=0.7$ is used. For all active facets (see below), EAF is set to one for both equations. This means that for both the water production and the recoil force, the mass sublimation rate is always 100\% active for all facets. Finally, the NGA produced by a region can be found via Eqn. \eqref{eq:FtoNGA}:
\begin{equation}\label{eq:FtoNGA}
    \vec{\mathbf{a}}_{NG}^{Region} = \frac{\sum_i F_i}{M_{67P}}~,
\end{equation}
by summing the facets $i$ forming a region on 67P. The mass taken for 67P equals $M_{67P}=9.982\cdot10^{12}$ from \cite{Patzold16}. 

As in \citet{attree-2019, attree-2024}, the thermal and force calculations are run 36 times over a comet rotation period and the summed curves averaged to make one data-point. This process is repeated at $\sim50-100$ points along the orbit for each comet, depending on the length of the observation arc, creating the water-production and NGF datasets.

\subsection{Application to Long Period Comets}\label{subsec:Application to LPC}
The thermophysical model is solved using the trajectories of C/2001 Q4, C/2008 A1, and C/2013 US10, along with the shape model of 67P. Using the Monte Carlo sampling technique of the 26 active regions of 67P, we can develop a family of possible realistic water production rates and their associated NGA curves. A sample of the active regions is considered valid based on the total water production compared to typical long-period comet secular brightness curves \citep{lacerda-2025}, followed by a comparison of the change in osculating elements caused by the sampled NGA against the observed changes in osculating elements in historic comets from the JPL SBDB. 

\subsubsection{Water production validation}
For each Monte Carlo sample of the active regions, the total water production is taken as:
\begin{equation}
    Q_{H_2O}^{Sampled}=\sum_n^kQ_{H_2O}^{Region}(n)~,
\end{equation} for a maximum of $k\in[5,26]$ regions.  The sum is compared to the water production converted secular brightness curves of historic comets. Historic brightness curves are converted to water production curves via Eqn. \eqref{eq:mag_to_Q} \citep{Jorda1991}:
\begin{align}\label{eq:mag_to_Q}
    Q_{tot}(r) = 10^{A-B\cdot m},\quad & Q_{tot}^l(r) = 10^{A-B\cdot m^l},& Q_{tot}^u(r) = 10^{A-B\cdot m^u}~,
\end{align} with $A=30.53$ and $B=0.234$ valid for LPCs at heliocentric distances (r) less than 3 AU from \citet{sosa-2011}. $m,~m^l$ and $m^u$ are the total median, lower and upper brightness magnitude over the total trajectory, which are determined via the total heliocentric magnitudes parameters at 1 AU as $m_1=8.28$, the median value, the lower trend $m_1^u=14.9$, and $m_1^l=2.5$, the upper trend, from Figure 12 of \citet{lacerda-2025}. These values have been chosen to capture the peak distribution. The brightening slope parameters are estimated from Figure 7 from \cite{lacerda-2025}, with the near-Sun slope (valid within 3.16 AU) as median $k_n=6.7285$, upper trend $k_n^u=10$ and lower trend $k_n^l=3$. A Monte Carlo sample is accepted if there exists a heliocentric distance
$r_c > 3$~AU such that $Q_{H_2O}^{Sampled}(r_c) = Q_{tot}^l(r_c)$ and
\[
Q_{tot}^l(r) \leq Q_{H_2O}^{Sampled}(r) \leq Q_{tot}^u(r)
\quad \forall \, r \in [r_c, q] ,
\]
where $q$ is the perihelion distance. Post-perihelion water production curve limitations are not accounted for. 

\begin{figure}
    \centering
    \includegraphics[width=1\linewidth]{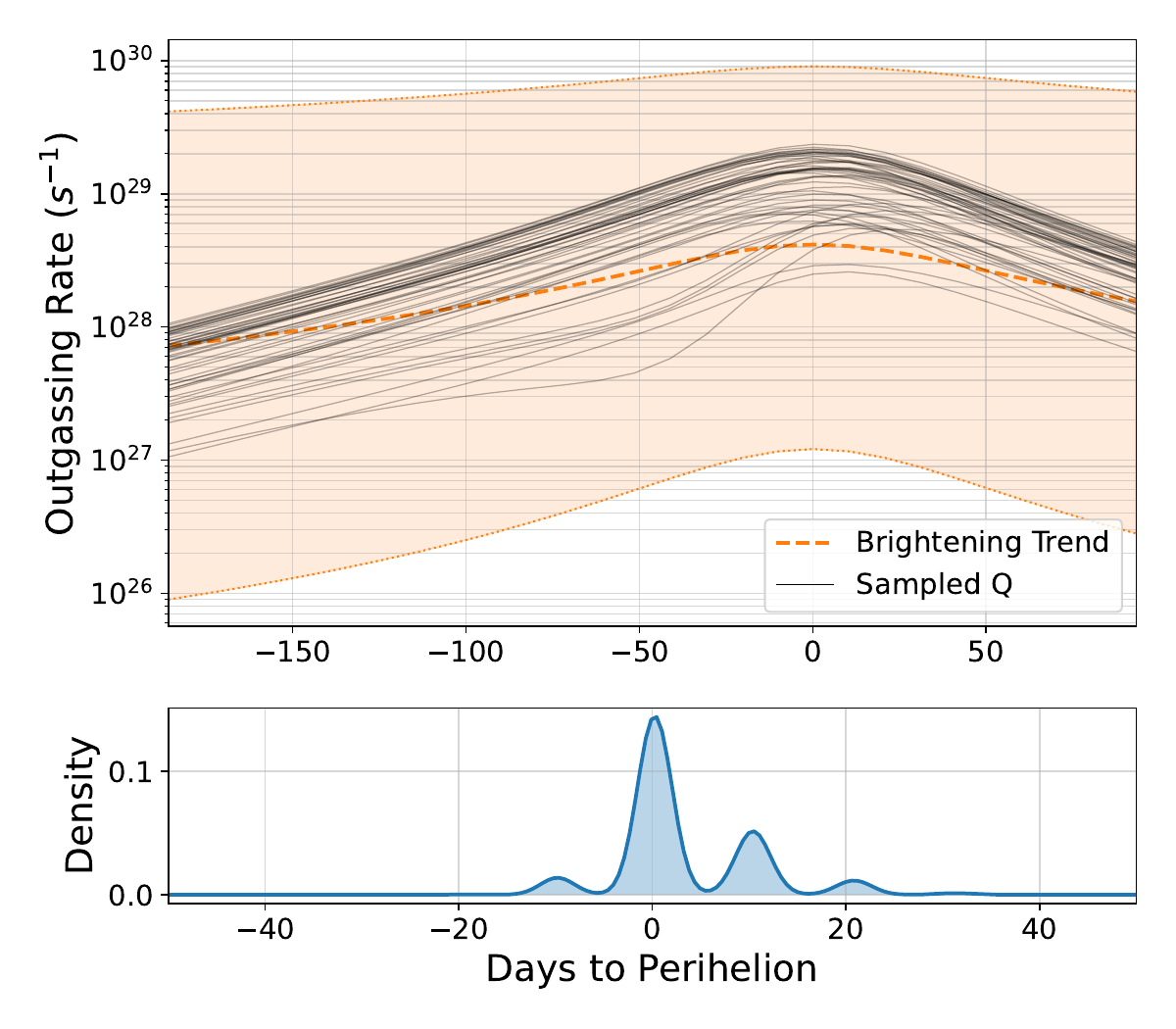}
    \caption{50 water production curves from the thermophysical model on the reference orbit of C/2001 Q4 (NEAT). The x-axis shows the time in days centred around perihelion, the y-axis shows the outgassing rate in mol/s. The subfigure shows the kernel density estimation for the peak outgassing times of all samples.} 
    \label{fig:QProduction_Q4}
\end{figure}

Figure \ref{fig:QProduction_Q4} (top) shows the total water production curves resulting from the 1000 Monte Carlo sampling of the active regions of 67P on the reference orbit C/2001 Q4 (NEAT). All Monte-Carlo samples of the active regions are valid, whilst for C/2008 A1 and C/2013 US10, one sample is invalid. This is a result of the wide range of absolute brightness magnitudes for DN LPC, possibly due to high activity levels \citep{lacerda-2025, sosa-2011}. The radius of 67P, which is only 2 km, is not adjusted, meaning that the higher envelope of the birghtness curves is not reached. In the subfigure, we can observe the kernel density estimation of the peak outgassing times. The majority of the water production samples peak slightly after perihelion, followed by $\approx10$ days post perihelion. All samples have peaked between 20 days before perihelion and 30 days after perihelion. 

\subsubsection{Change in osculating elements validation}
The NGA accelerations of the valid samples are summed via Eqn.~\eqref{eq:sum_NGA}
\begin{equation}\label{eq:sum_NGA}
    \vec{\mathbf{a}}_{NG}^{Sampled} = \sum_n^k \vec{\mathbf{a}}_{NG}^{Region}(n)~,
\end{equation} over the same k regions. This results in a sparse time acceleration vector in the $RTN$ frame in $AU/day^{2}$. When passing this vector to the numerical integrators, the thermophysical accelerations are interpolated using an m=8 ($7^{th}$ order) Lagrange interpolator due to the coarse timestep used in the thermophysical model.

The change in osculating elements resulting from such an acceleration curve is computed as the difference between the initial state at the first observed epoch and the final propagated state at the last observed epoch, where the propagation is performed using the open-source \textit{REBOUND} code\footnote{https://rebound.hanno-rein.de/} \citep{rein-2011}. The initial osculating states are obtained from the JPL SBDB for the three reference comets, and each is modelled as a test particle in the heliocentric Ecliptic J2000 reference frame. To isolate the effects of the NGAs, for these propagations only, the dynamical model is a two-body Sun-comet problem with the thermophysical accelerations applied and interpolated using spline methods through \textit{REBOUND}x extension package\footnote {https://reboundx.readthedocs.io/en/latest/}. Integration is carried out with \textit{TRACE} \citep{lu-2024}.

\begin{figure*}
    \centering
    \includegraphics[width=1\linewidth]{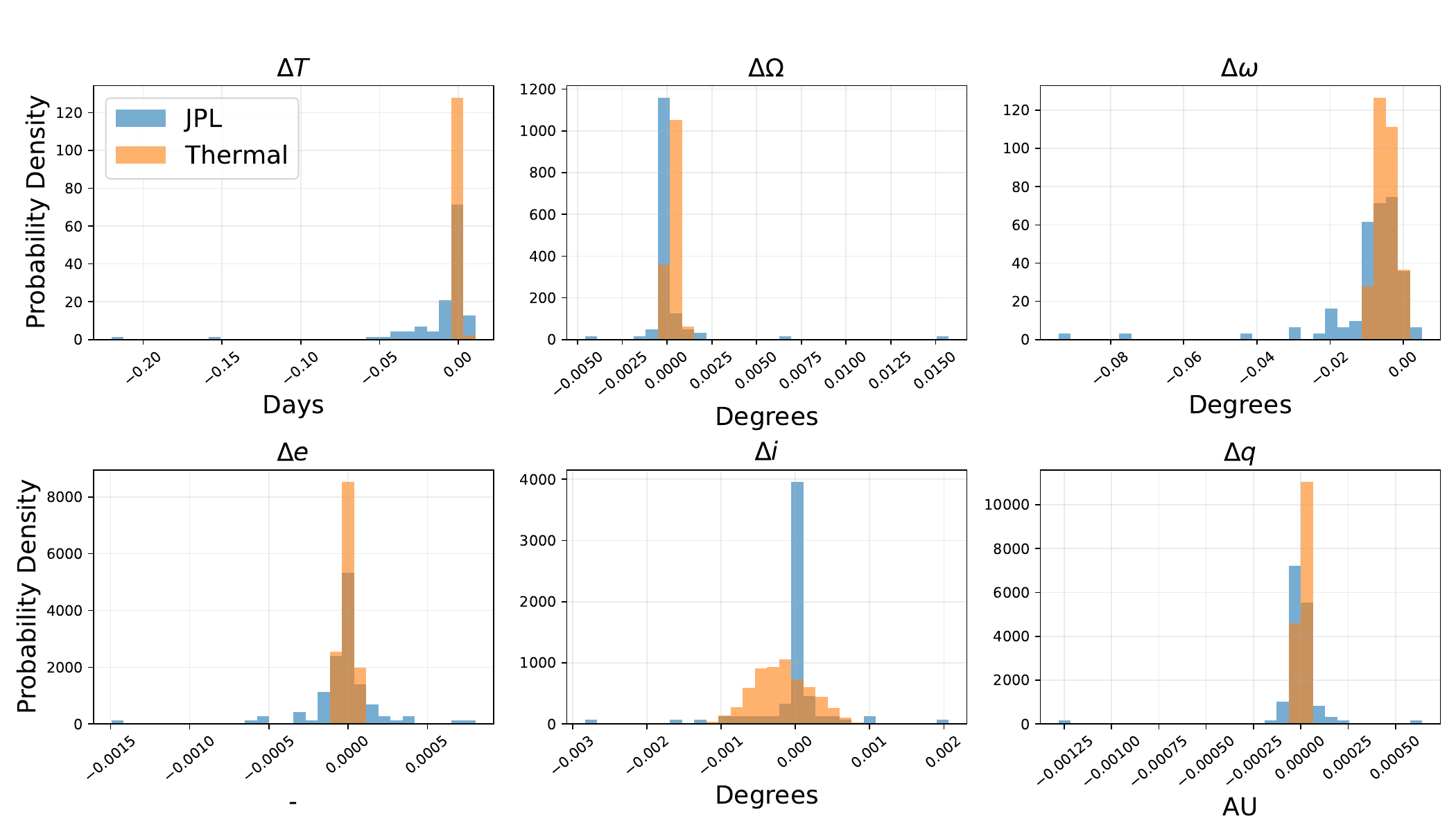}
    \caption{Distributions of changes in osculating elements of the 94 Marsden LPC solutions available from JPL compared to the 1000 Monte-Carlo sampled thermophysical models on reference orbit C/2001 Q4 (NEAT). A force multiplication of 4 is used. Bin-sizes from left to right and top to bottom are: 0.00769 d, 0.000677 $^{\circ}$, 0.00331 $^{\circ}$, 7.652$\times10^{-5}$, 0.000163 $^{\circ}$, and 6.403$\times10^{-5}$ AU, respectively.}
    \label{fig:DeltaOsculatingQ4}
\end{figure*}

A reference distribution for the change in osculating elements is generated using the same propagation settings. Comets are selected via the SBDB API filter for long-period comets classified as Hyperbolic, Parabolic, or Comet types with at least the $A_1$ Marsden parameter available. The reference set is constructed using both symmetric and asymmetric Marsden $H_2O$ and $CO$ NGA models. A total of 94 comets pass the filter requirements to generate the test distribution. A Monte-Carlo sample of the thermophysical model (Model II) is valid if the change in all 6 osculating elements falls within the $3\sigma$ distribution of the reference Marsden distribution. 

Initial thermophysical perturbed distributions show narrow peaks near zero for all six osculating elements, indicating the distributions do not reflect the observed behaviour of real comets. The thermophysical distribution is adjusted to account for the 2 km radius and 500 kg/m$^3$ density of 67P by scaling up the retrieved accelerations from the thermophysical model. Considering the possibility that LPCs could be small and hyperactive (e.g. half the radius of 67P, resulting in twice the area:mass ratio) and/or low density (e.g. half the density of 67P) \citep{sosa-2011}, scaling of 2x the force and 4x the force are applied. Figure \ref{fig:DeltaOsculatingQ4} shows the distribution of the six osculating elements for the 94 Marsden simulated comets and 1000 Monte-Carlo samples of the thermophysical model with a force multiplication of 4 for C/2001 Q4 (NEAT). We observe that for the osculating elements $\Delta\omega$, $\Delta e$, $\Delta q$, and $\Delta T$, the peaks of the reference distribution are well captured by the thermophysical model (Model II), but the full distribution of the reference NGAs is not filled out. Model II shows a well-sampled distribution for $\Delta i$, but does not capture the peak around zero. $\Delta\Omega$ shows a positive-sided distribution and is not centred around the Marsden peak. Similar distributions are found for C/2013 US10, where the $\Delta\Omega$ is negatively distributed, and C/2008 A1, where it is centred around the Marsden peak. These differences can be explained by the difficulty in determining $A_{3}$ values in orbital fits, where only $45\%$ of  the reference sample have a non-zero $A_{3}$. As shown by \citet{Sekanina1993}'s equations 11, $\Delta\Omega$ and $\Delta i$ are automatically equal to zero for Marsden models with $A_{3}=0$, explaining the bias in the JPL SBDB dataset.

\begin{figure*}
    \centering
    \includegraphics[width=1\linewidth]{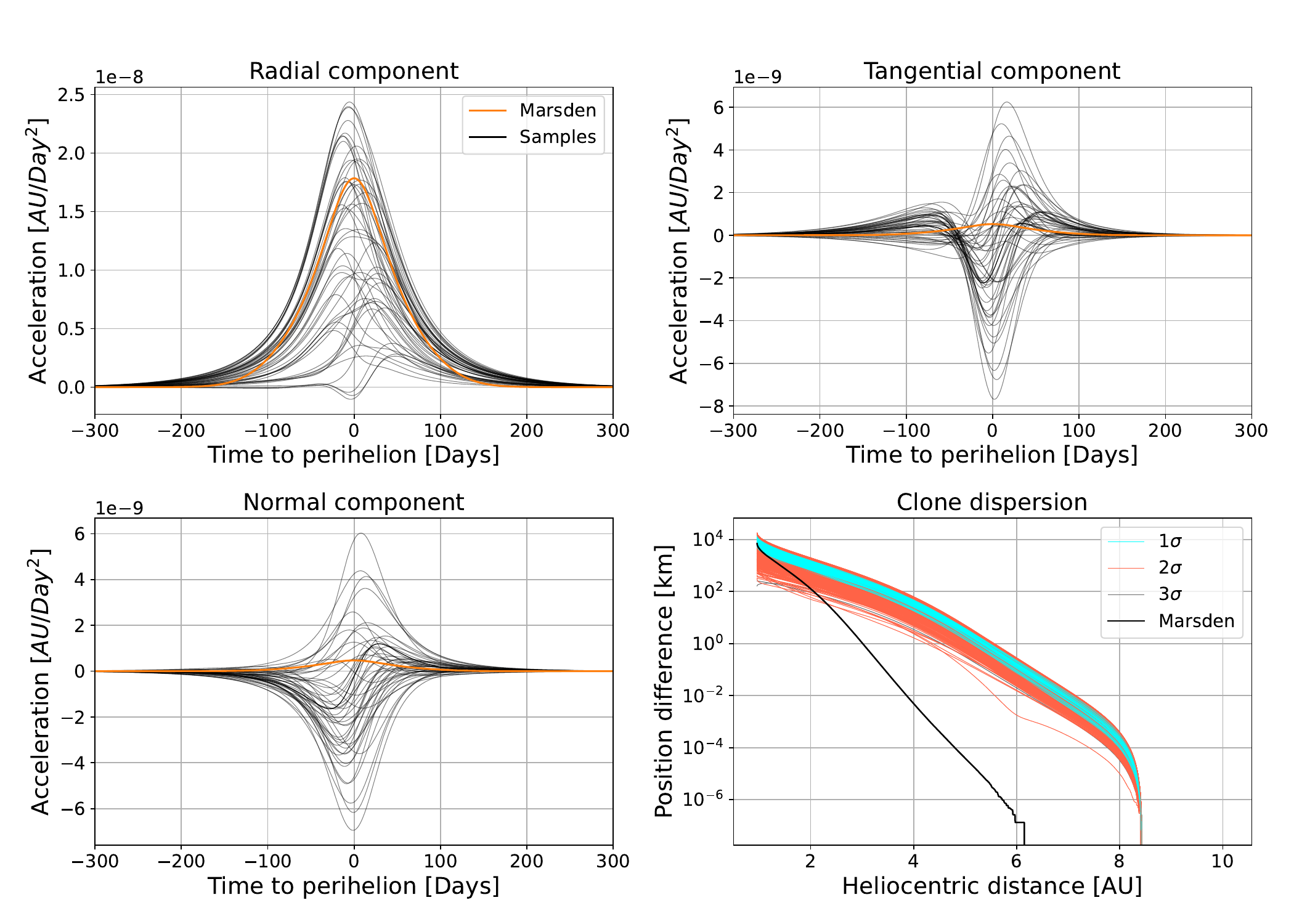}
    \caption{Top left: 50 sampled radial accelerations, top right: 50 sampled tangential component and bottom left: 50 sampled normal component of the Monte-Carlo sampled thermophysical model on the reference trajectory of C/2001 Q4 (NEAT) with Marsden as comparison. The bottom right shows the trajectory deviation compared to a two-body problem for the 1000 Monte Carlo samples and the Marsden result. The $\sigma$'s relate to the distribution at perihelion.}
    \label{fig:RTN_DIV_Q4}
\end{figure*}

Some example sampled Model II accelerations are shown in Figure \ref{fig:RTN_DIV_Q4}, and represent the same sampled regions as the water production curves in Figure \ref{fig:QProduction_Q4}. We observe a wide range of possible acceleration curve shapes compared to the Marsden solution (Model I) for this comet (C/2001 Q4). In the radial component, we can note the asymmetry in the acceleration curves, with some having two local peaks, one pre- and one post- perihelion, as well as occasional negative peaks. The "spot model" of \cite{Sekanina1993} demonstrated that such behaviour can arise for specific combinations of active regions and spin-axis orientation. In the tangential and Normal components, we observe sign changes leading up to- and/or after the perihelion time. Such sign changes are not captured by the \cite{marsden-1973} formulation. Also notable, in the lower right of Figure \ref{fig:RTN_DIV_Q4}, is that significant divergence (100s - 1000s km) in the trajectories relative to a purely gravitational orbit can occur at larger heliocentric distances in the thermophysical models than in the Marsden model.

\section{Results}
\label{sec:results}

\subsection{Model I: Marsden-Formulated NGA Trajectory Uncertainty}
\label{Results:Masden}
In this subsection, the trajectory uncertainty of a Marsden-based truth trajectory (Model I) is evaluated for the three Comet Interceptor virtual targets. Using 1000 Monte Carlo samples drawn from the covariance matrix produced by the Case A and Case B estimators, the uncertainty is assessed at the time of perihelion. Case C is not evaluated  as it is not applicable for purely symmetric Marsden truth orbits. Figures \ref{fig:Marsden_2001Q4}, \ref{fig:Marsden_2008A1}, and \ref{fig:Marsden_2013US10} present the results for the Case A and Case B estimators for the comets C/2001 Q4, C/2008 A1, and C/2013 US10, where the points show the median and interquartile range of the true trajectory error, $\Delta$, according to Eqn.~\ref{eq:Delta}.

\subsubsection{Case A – Pure Gravity}
For all three comets, the pure-gravity orbit fit (blue points), applied to observations perturbed by Marsden NGAs (Case A, Model I), fails to recover the true trajectory. Across all three, the median position error, $\tilde{\Delta}$, stabilises after the second observation arc and remains approximately constant thereafter. This is despite the associated uncertainties (i.e.~the spread in $\Delta$ values, here quantified by interquartile range) decreasing, indicating that the formal errors from the POD are significantly underestimating the true uncertainty, due to the presence of the unmodelled accelerations. This is confirmed by visual inspection of the post-fit residuals in right-ascension and declination, which are not flat with time, and still show significant trends. One day prior to perihelion, the effects of the NGA are still partially absorbed into the estimated initial state. The estimated median errors compared to the reference orbits are $\tilde{\Delta}=1094^{+4.9}_{-4.5}$ km for C/2001 Q4, $7795^{+2.9}_{-2.9}$ km for C/2008 A1 and $1821^{+3.8}_{-3.8}$ km for C/2013 US10.

Clearly, a pure-gravity fit is unable to reproduce the comets' full, perturbed trajectories, but it might still be expected that an osculating orbit could be found that would reproduce a short orbital arc near perihelion with an accuracy sufficient to decrease the observed error at intercept. To this end, we also attempted pure-gravity fits to 0.5 AU-long arcs of observations, working backwards from time of interception, in a similar technique to that used when planning occultations \citep{Desmars2015}. We were unable, however, to improve the accuracy of the solutions using this method. This is likely due to the long arcs (0.5 AU, relevant for predictions at 60 days) incorporating too much NGA perturbation compared to the short arcs of a day, or a few days, used in occultation studies \citep{Desmars2015}, and in improving trajectories between the Vega I, Vega II, and Giotto encounters with comet Halley \citep{Muench1986}.

The results for the velocity error at perihelion are shown in the figures in Appendix \ref{sec:appendix_velocities}, and show a slightly better accuracy, with respect to the mission goal of 1 m s$^{-1}$ than for position, but still generally fail to reach this accuracy sixty days before intercept.

\subsubsection{Case B - Marsden $A_i$ }
We now describe the pre-perihelion fits of symmetric Marsden trajectories to the underlying symmetric Marsden truth orbits (Model I, Case B: orange triangles in Figs.~\ref{fig:Marsden_2001Q4}, \ref{fig:Marsden_2008A1}, and \ref{fig:Marsden_2013US10}). The observation cadence for estimating the $A_i$ parameters differs slightly because the estimation process requires observations in the NGA regime of the trajectory. Therefore, observations are removed 20 at a time per POD until one of the estimated $A_i$ accelerations exceeds $1\times10^{-5}$ m s$^{-2}$.

We can observe across all three figures that close to perihelion, the position error is reduced compared to the pure gravity fit. For C/2001 Q4, we find a final position error of $\tilde{\Delta}=53^{+13}_{-13}$ km at the last observation, one day before perihelion with a total of 996 observations. Sixty days before perihelion, we have an error of $2868^{+1574}_{-1376}$ km. The target requirement of 1000 kilometres uncertainty is not satisfied until forty days before perihelion with an error of $452^{+298}_{-201}$ km. Similarly, for C/2008 A1 sixty days before perihelion, position error is $4250^{+1767}_{-1773}$ km, and the target requirement is not met until forty days, with $938^{+469}_{-344}$ km. Finally, C/2013 US10 has an error at the last observation 1 day before perihelion of $393^{+262}_{-176}$ km, while at sixty days the position error of $4415^{+1921}_{-1832}$ km also does not satisfy the mission goal.

Overall then, modelling incoming comet NGAs with a symmetric Marsden model (a reminder that none of these three comets have a fitted $\tau$ term), and then attempting to recover those trajectories with a Marsden fit, improves accuracy compared to a pure-gravity fit. However, astrometric uncertainties and difficulties in the fit are still enough that the position error at 60 days is of the order of several-, not one-, thousand kilometres. Further, some of the 'early' fits with the Marsden model, before the 60 day mark, actually produce worse accuracy at perihelion than pure-gravity fits. Clearly the Marsden model, which typically fits orbitally integrated accelerations over several orbits, struggles when applied to orbital arcs containing little NGA information (see the times at which NGAs `turn on' in Fig.~\ref{fig:RTN_DIV_Q4}), and care must be taken when orbit fitting here.

\subsection{Model II: Thermal-Formulated NGA Trajectory Uncertainty}
In this subsection, the trajectory uncertainty of a Thermophysical-based truth trajectory (Model II) is evaluated. 500 Monte Carlo samples are drawn from the covariance matrix produced by the Case A, Case B and Case C estimators for each of 1000 unique acceleration curves. The uncertainty is assessed at the time of perihelion and presented as before. Figures \ref{fig:Thermal_2001Q4}, \ref{fig:Thermal_2008A1}, and \ref{fig:Thermal_2013US10} show these results for the comets C/2001 Q4, C/2008 A1, and C/2013 US10. Velocity results are again shown in Appendix \ref{sec:appendix_velocities}.

\subsubsection{Case A}
\begin{figure}
    \centering
    \includegraphics[width=1\linewidth]{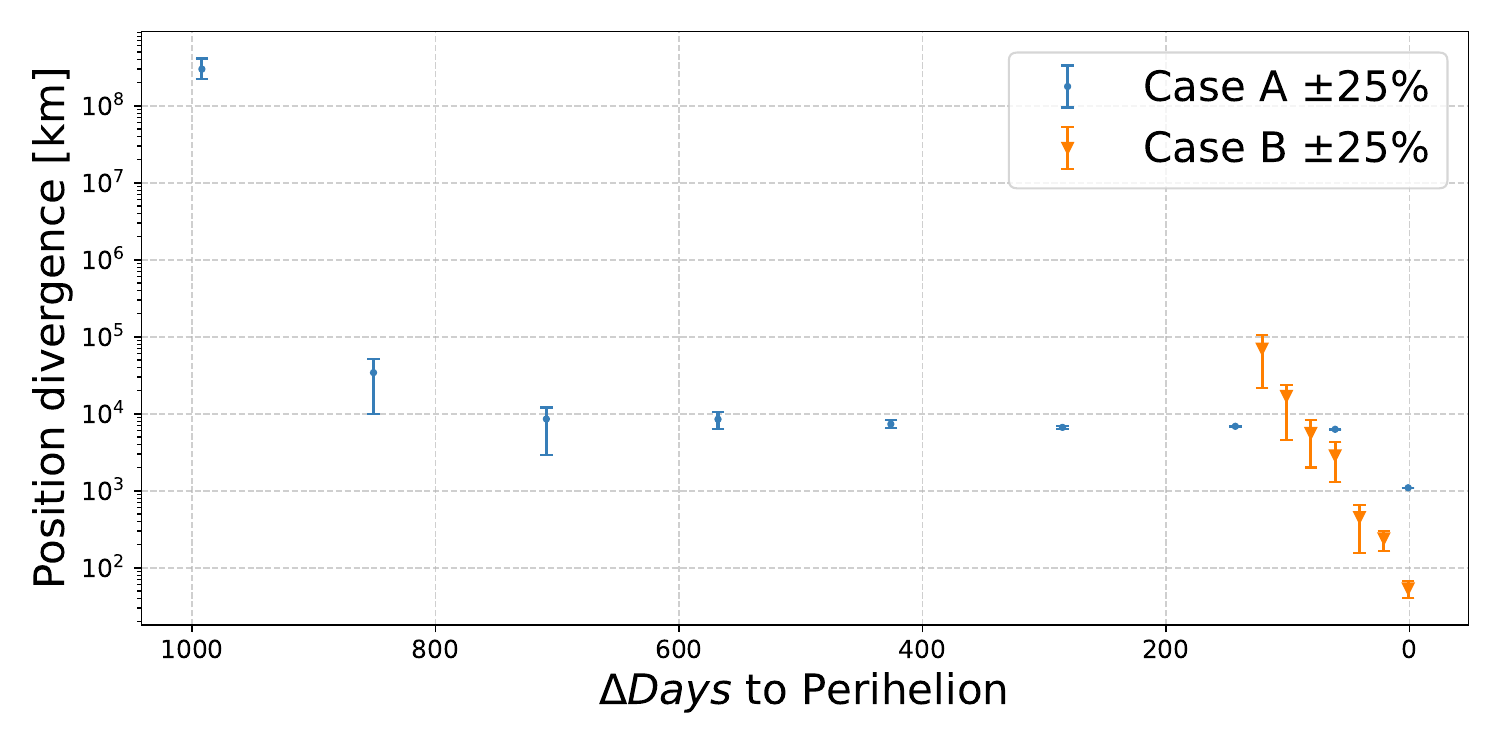}
    \caption{Perihelion position distributions, $\Delta$, of fitted comet C/2001 Q4 orbits compared to a Marsden reference orbit (Model I). The x-axis describes when the last observation took place, the y-axis describes the distribution at perihelion, and the points show medians with error bars given by the interquartile range. Blue dots are Case A and orange triangles, Case B.}
    \label{fig:Marsden_2001Q4}
\end{figure}

\begin{figure}
    \centering
    \includegraphics[width=1\linewidth]{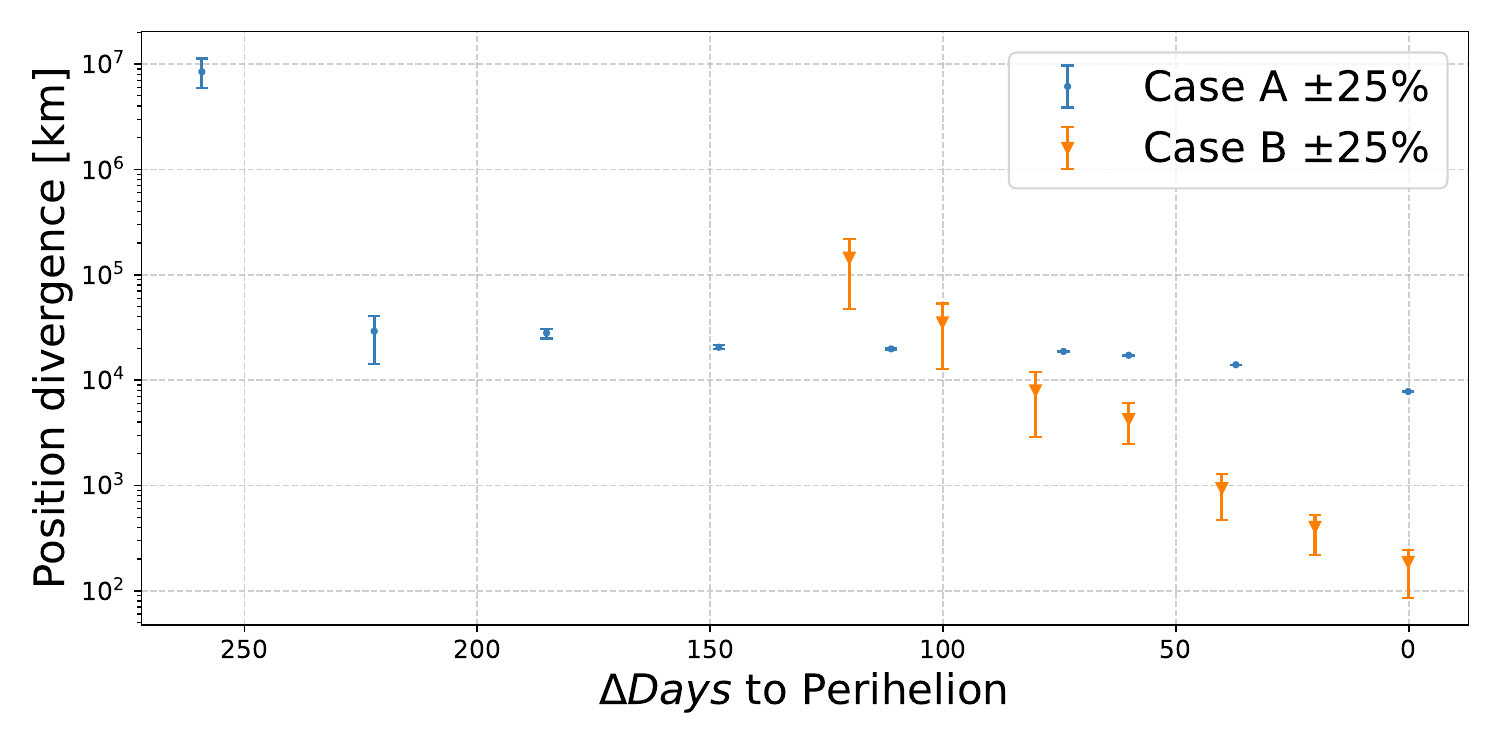}
    \caption{Perihelion position distributions, $\Delta$, of fitted comet C/2008 A1 orbits compared to a Marsden reference orbit (Model I). The points and axes are as above in Fig.~\ref{fig:Marsden_2001Q4}.}
    \label{fig:Marsden_2008A1}
\end{figure}

\begin{figure}
    \centering
    \includegraphics[width=1\linewidth]{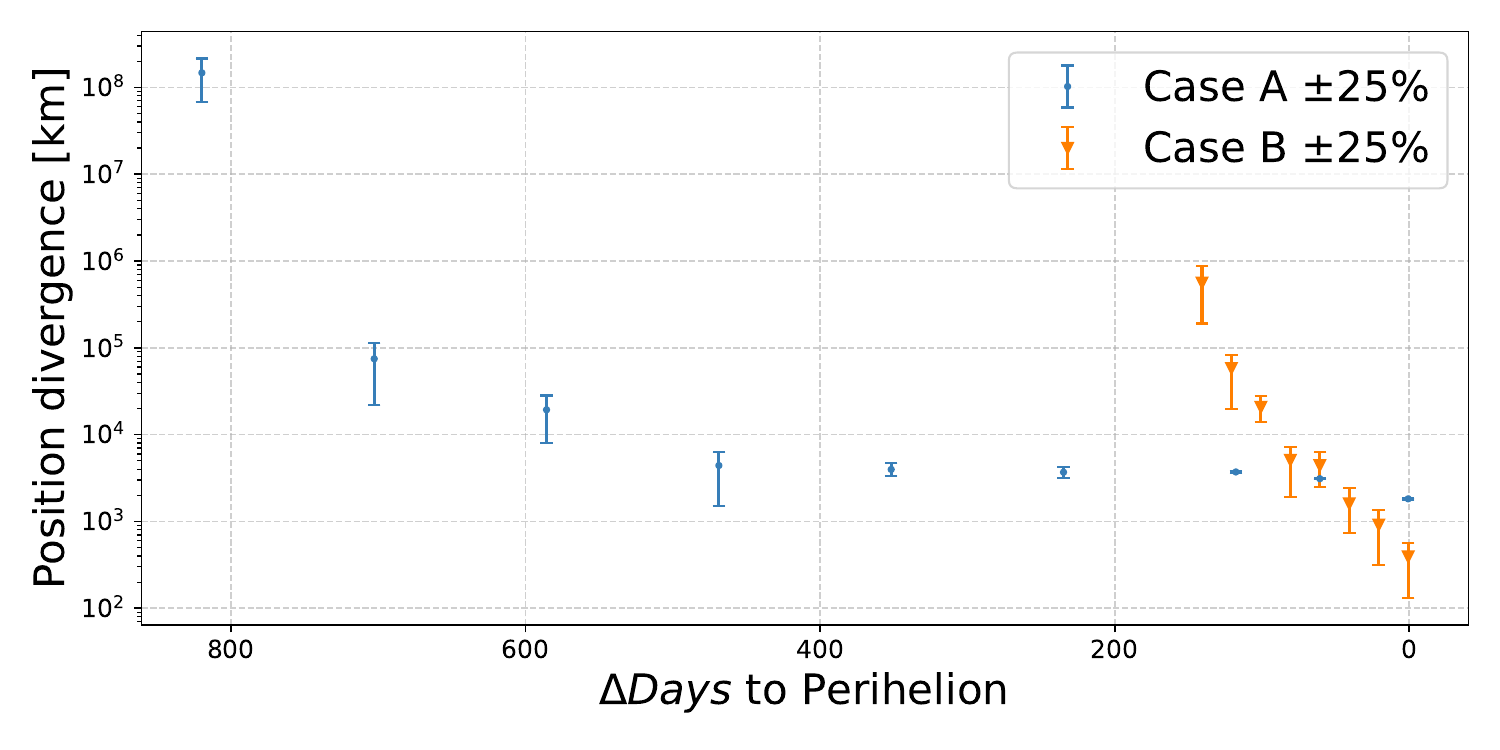}
    \caption{Perihelion position distributions, $\Delta$, of fitted comet C/2013 US10 orbits compared to a Marsden reference orbit (Model I). The points and axes are as above in Fig.~\ref{fig:Marsden_2001Q4}.}
    \label{fig:Marsden_2013US10}
\end{figure}

Pure gravity fits to the thermophysical acceleration curves (Model II, Case A) show the same general trends as when fitted to Marsden-style NGAs; i.e.~the fitted orbits are unable to reproduce the reference trajectory to better than $\sim1000$ km accuracy a few days before perihelion, or $\sim10,000$ km around sixty days before. Error bars in Figs.~\ref{fig:Thermal_2001Q4}, \ref{fig:Thermal_2008A1}, and \ref{fig:Thermal_2013US10} are larger than in Figs.~\ref{fig:Marsden_2001Q4}, \ref{fig:Marsden_2008A1}, and \ref{fig:Marsden_2013US10} because they now show the interquartile range over the 1000 sampled Model II truth trajectories, rather than the single Marsden-style trajectory of above.

\subsubsection{Case B}
Similarly to above, the addition of a symmetric Marsden model with $A_{i}$ parameters improves the accuracy of the fitted orbits, particularly within the last $\sim60$ days when the NGAs are largest (Fig.~\ref{fig:RTN_DIV_Q4}). Before that, when the NGAs are small, the fitting procedure struggles to incorporate them, often resulting in worse fits than the pure gravity Case A, in the same way as for the Marsden reference trajectories. This demonstrates again the difficulty in predicting `real' thermophysical accelerations curves around perihelion by extrapolating data from larger heliocentric distances.

Again, comet C/2001 Q4 has the best final trajectory error of $318^{+130}_{-105}$ km, and $6800^{+3662}_{-1808}$ km sixty days before. Notably, both of these are significantly worse than the previous section's Marsden fits, due to both the greater difficulty in fitting the more complex thermophysical acceleration curves, and the increased scatter in fitting 1000 separate curves rather than a single Marsden trajectory. The latter error is again outside the required 1000 km. For the other two comets, the accuracies at perihelion and at 60 days before are even worse, with Case B fits struggling to reproduce the reference trajectories to within 1000 km, even a few days before. As shown in figure \ref{fig:RTN_DIV_Q4}, many of the thermophysical acceleration curves are asymmetric, suggesting an asymmetric NGA model may be necessary. This is addressed in Case C below.

\subsubsection{Case C}

When introducing asymmetry into the NGA model, our optimisation routine encountered significant difficulties, with many LSQ-fits resulting in ill-defined covariance matrices or failing to converge completely. This is due to the well-known instability caused by introducing the time-offset, $\tau$ (most historical comets are fitted with $\tau=0$), especially when considering the limited orbital arc. We addressed this by introducing an \textit{a priori} constraint on the covariance matrix to help guide the fitting routine to physically realistic values. We imposed \textit{a priori} one-sigma uncertainties of 1000 km and 10 m s$^{-1}$ on the initial position and velocity (state vector), and $1\times10^{-5}$ m s$^{-2}$ and 50 days, on $A_{i}$ and $\tau$, respectively. The effects of applying these \textit{a priori\it} constraints were also tested with Case B, with almost identical results (e.g.~$\tilde{\Delta}=6800^{+3662}_{-1808}$ km at sixty days from perihelion for Q4, above, changed to $6802^{+4044}_{-1726}$ km, with the \textit{a priori} constraints). Additionally, we continued the orbital integrations past perihelion and performed one final Case C fit to the full orbital arc, using the same observation cadence and weighting, in order to asses the limitations of relying on pre-perihelion data only, versus the full arc.

Even given these constraints, some fits still failed to converge, leading us to discard them in the final results. For some comets and fits with the shortest observing arcs, up to $\sim50\%$ of the acceleration curves are discarded, while for the last pre-perihelion point, over $83\%$ of sampled NGA curves are successfully fit. For the full orbital arcs, a solution is found for all samples. The differing covariance treatments and number of accepted fits means that, strictly speaking, the statistics of the Case C results are not directly comparable with those of the other cases. However, the medians and interquartile ranges still give a good sense of the average improvement over Cases A and B. The failure to find Case C solutions for some sampled accelerations, meanwhile, reflects real limitations in the ability of the asymmetric Marsden model to fit physical acceleration curves with limited data, that may well occur with the true Comet Interceptor target. This may necessitate a falling-back to the symmetric case at times. When comparing the failed fits against the successfully fitted samples, we find differing distributions in their underlying acceleration curves. These differences vary in a non-systematic way between the three comets, demonstrating that the limiting-factor in determining their orbits (such as total NGA magnitude, acceleration vector direction, or peaking time) depends on the observing arc and specific orbital geometry in a complex way. Analysis of the underlying accelerations curves will be discussed further in Sect.~\ref{Discussion:fittedparameters}, below.

\begin{figure}
    \centering
    \includegraphics[width=1\linewidth]{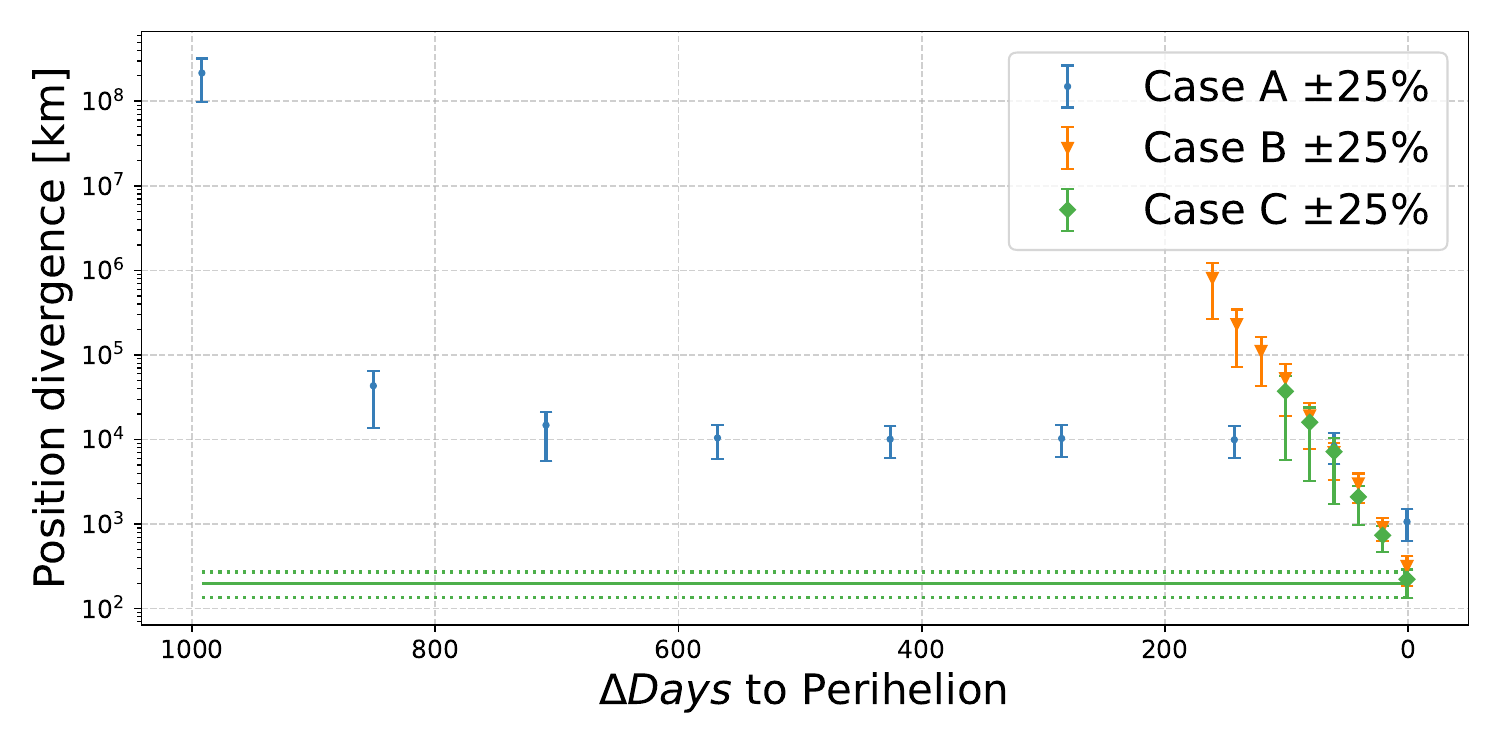}
    \caption{Perihelion position distributions, $\Delta$, of fitted comet C/2001 Q4 orbits compared to thermophysically-perturbed reference orbits (Model II). The x-axis describes when the last observation took place, the y-axis describes the distribution at perihelion, and the points show medians over the full suite of Monte Carlo NGA samples with error bars given by the interquartile range. Blue dots are Case A, orange triangles, Case B, and green diamonds, Case C. The horizontal solid and dotted lines denote the full-orbit Case C median and interquartile range respectively.}
    \label{fig:Thermal_2001Q4}
\end{figure}

\begin{figure}
    \centering
    \includegraphics[width=1\linewidth]{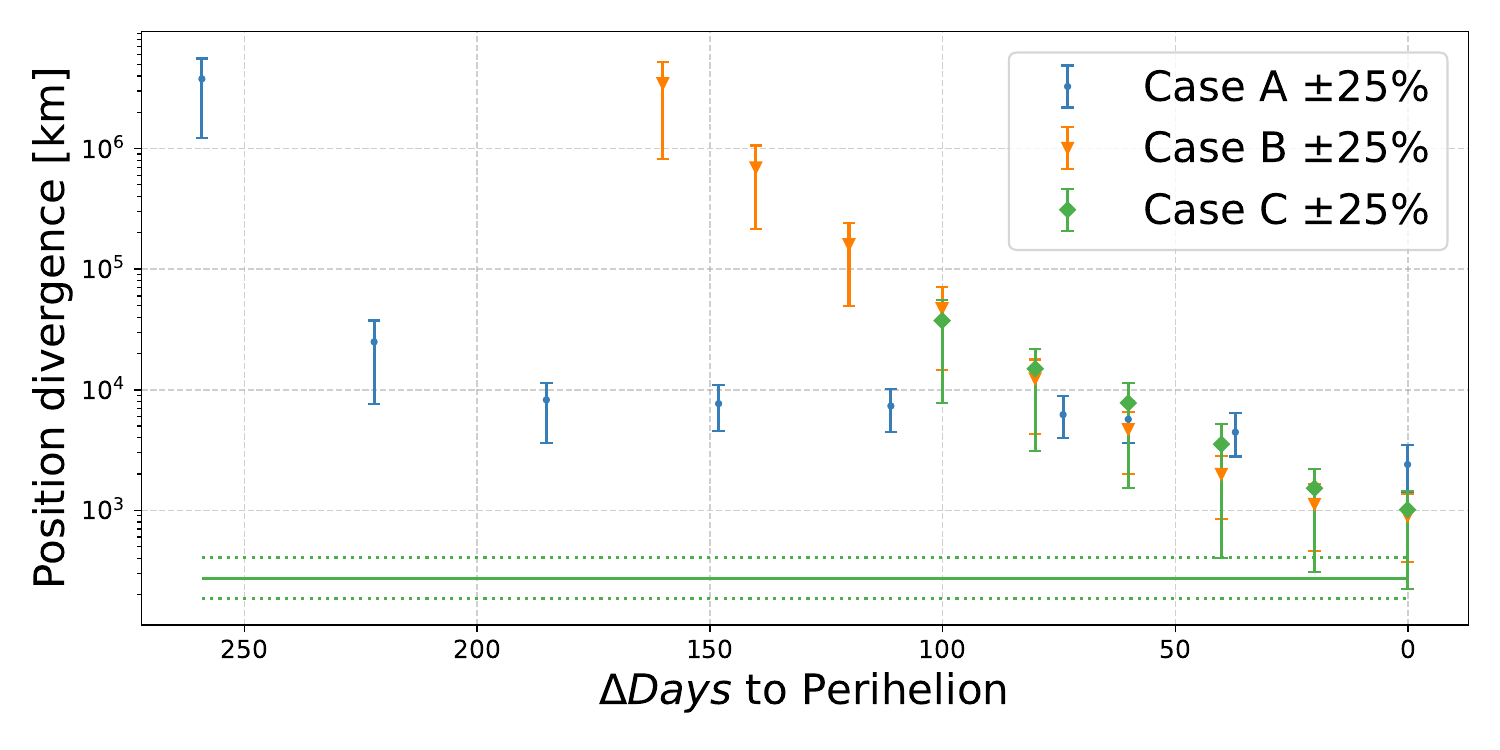}
    \caption{Perihelion position distribution, $\Delta$, of fitted comet C/2008 A1 orbits compared to thermophysically-perturbed reference orbits (Model II). The points and axes are as above in Fig.~\ref{fig:Thermal_2001Q4}.}
    \label{fig:Thermal_2008A1}
\end{figure}

\begin{figure}
    \centering
    \includegraphics[width=1\linewidth]{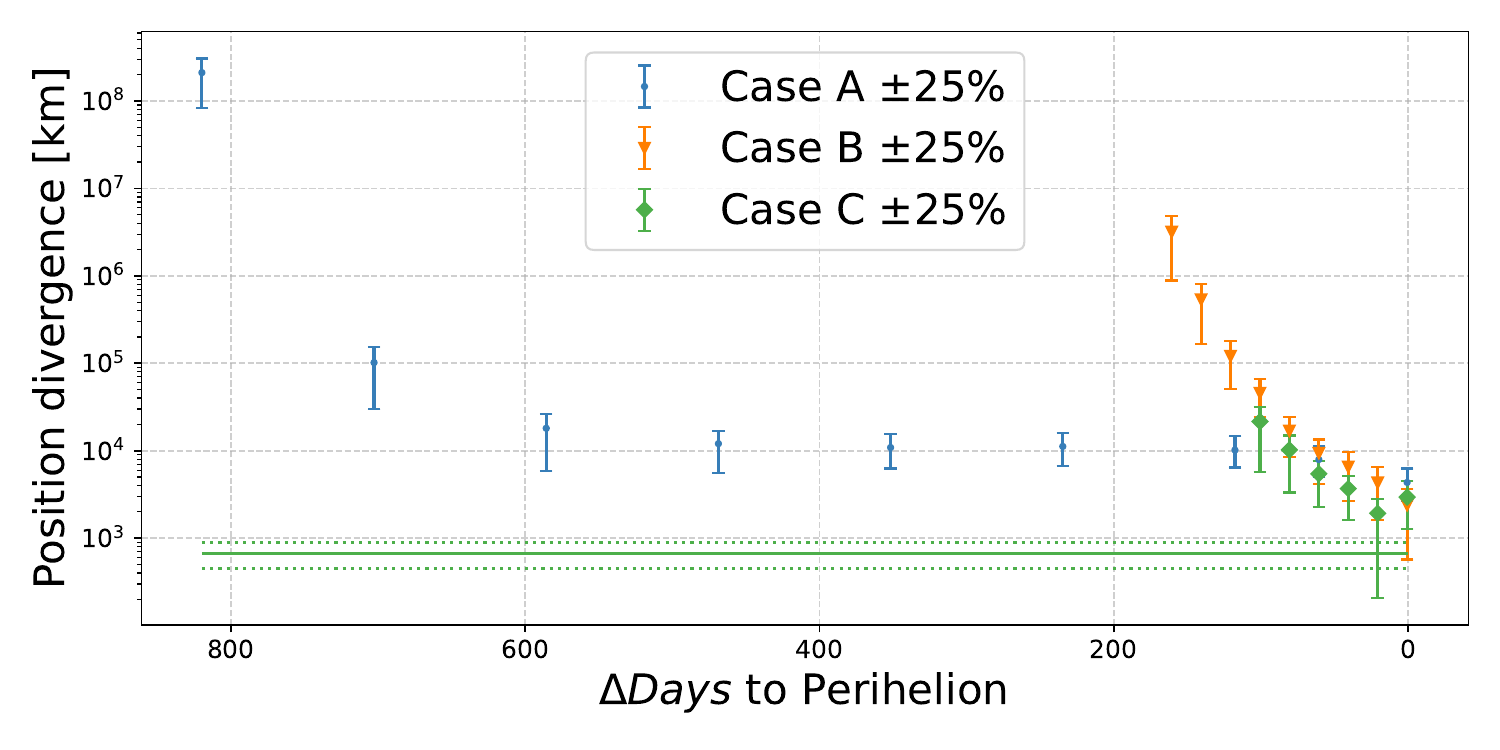}
    \caption{Perihelion position distribution, $\Delta$, of fitted comet C/2013 US10 orbits compared to thermophysically-perturbed reference orbits (Model II). The points and axes are as above in Fig.~\ref{fig:Thermal_2001Q4}.}   
    \label{fig:Thermal_2013US10}
\end{figure}

Figs.~\ref{fig:Thermal_2001Q4}, \ref{fig:Thermal_2008A1}, and \ref{fig:Thermal_2013US10} show that, when fitting the full orbital arc (full-arc $\tilde{\Delta}$ denoted by horizontal solid lines with dotted lines showing the interquartile range), sub-one-thousand kilometre accuracy at perihelion is assured for all the studied comets. Compared to this, the fits up to perihelion vary by comet, with C/2001 Q4 able to be fitted with an accuracy approaching the full-arc solution even with only pre-perihelion data, due to its long pre-perihelion observing arc and close Earth approach here (see Fig.~\ref{fig:rE}). This is not the case for C/2008 A1 or C/2013 US10, who's orbits are significantly improved when including post-perihelion data. The latter comet actually experiences a decline in accuracy in the last point pre-perihelion as compared to twenty days before, due to its quickly increasing Earth-centric distance here reducing the accuracy of these observations (Fig.~\ref{fig:rE}). Interestingly, the full orbital solution improves the situation again, demonstrating the importance of both Earth-centric distance and a consideration of all the data. For the former comet (C/2008 A1), the relative smaller number of observations leads to a lower accuracy overall and sometimes leads to a worse performance for the Case C fits than for Case B. Similarly to above, asymmetric Marsden models can perform worse than pure-gravity fits far from perihelion when the NGAs are small.

Overall, the median position error at sixty days remain at several thousand kilometres, with interquartile ranges extending to ten-or-more thousand kilometres, for all three comets, and for all three orbit fitting cases. This demonstrates the difficulty in predicting cometary positions around perihelion at times and distances significantly before that, when NGAs are just beginning to have an influence on the trajectories.


\subsection{Effects of increased astrometric noise}
\label{Results:astrometry}

All of the orbit determinations above rely on an assumed LSST performance of 0.1 arcseconds for the astrometric precision of the synthetic observations. Final astrometric precision may be significantly worse than this if, for example LSST's operating performance is not as good as expected, or it cannot observe the target comet due to it being in the northern sky and other observatories must be relied on. To test the effects of this on our POD, we re-ran the Model II, Case C solutions with an astrometric noise of 0.5 arcseconds, corresponding to a rough average-precision of Minor Planet Centre contributing observatories \citep{Veres2017}.

The results are shown below in Appendix \ref{sec:appendix_noise}, figures \ref{fig:Thermal_2001Q4_05}, \ref{fig:Thermal_2008A1_05}, and \ref{fig:Thermal_2013US10_05}. Median errors from the thermophysically modelled orbits, $\tilde\Delta$, are larger at all times and for all comets than for the case with 0.1 arcsecond astrometry. Perihelion position errors from sixty days are generally several tens of thousand kilometres with this level of astrometric accuracy.

Differences between the final accuracy at different noise levels are minimised in the full-arc solutions, as compared to the incoming orbit points, demonstrating that increased number of observations and, more importantly, arc-length can compensate somewhat for the quality of said observations. For the 'well-behaved' comet C/2001 Q4, the last point pre-perihelion based on high astrometric accuracy observations approaches the $\tilde\Delta$ of the full-arc solution with low astrometric accuracy. This demonstrates clearly that both observing arc-length and observing accuracy are important in predicting cometary positions.

\section{Discussion}
\label{sec:discussion}

The following discussion is framed in terms of the Comet Interceptor navigation requirements, but the results also illustrate more general principles governing the predictability of dynamically new comet trajectories under realistic non-gravitational perturbations. In particular, the framework separates uncertainties arising from observation geometry from those arising from imperfect physical modelling, making it applicable to a broader range of orbit-determination problems.

\subsection{Analysis of the Fitted Acceleration Parameters}
\label{Discussion:fittedparameters}

Before moving on to discuss the trajectory uncertainty results in terms of the mission requirements, we first pause to investigate the physical interpretability of the fitted accelerations terms in terms of our underlying Model II thermophysical NGA curves.

\begin{figure}
\resizebox{\hsize}{!}{\includegraphics[scale=0.5]{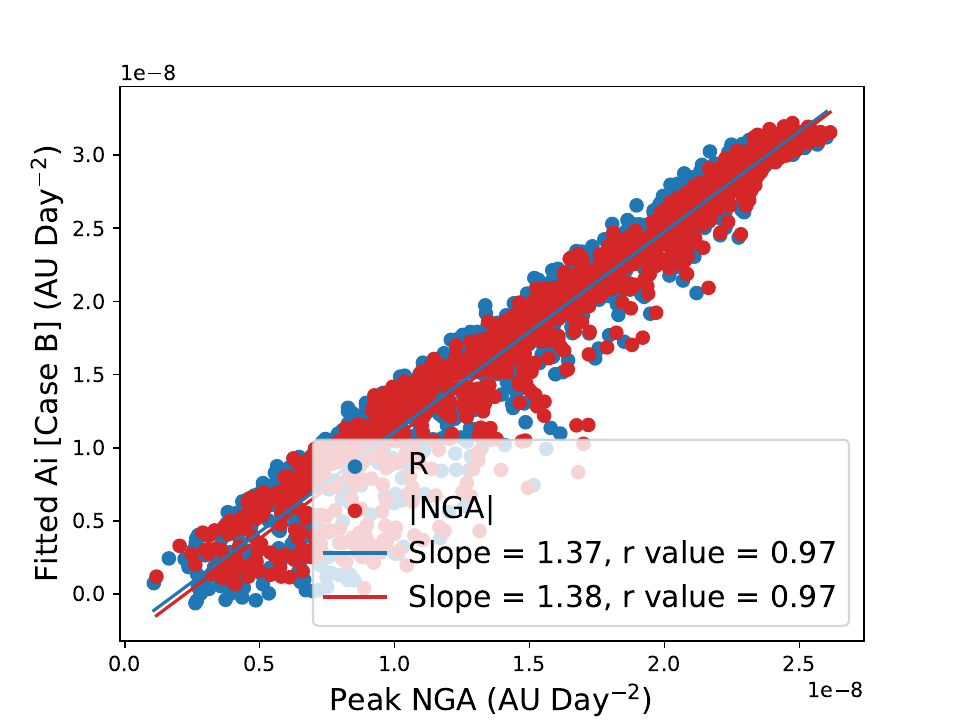}}%
\qquad
{\includegraphics[scale=0.5]{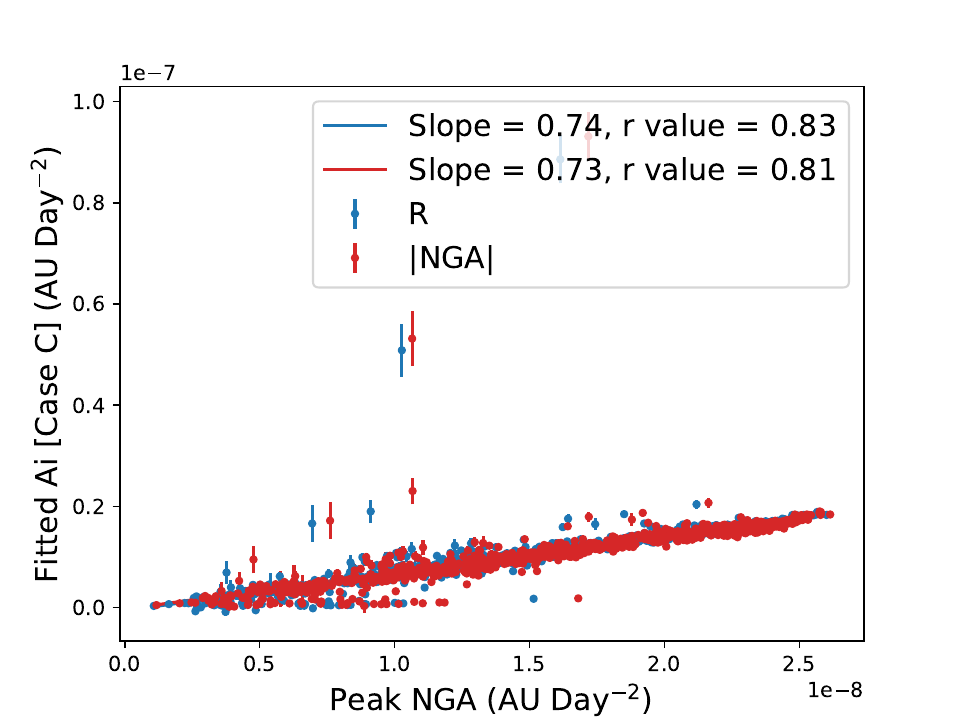}}%
\qquad
{\includegraphics[scale=0.5]{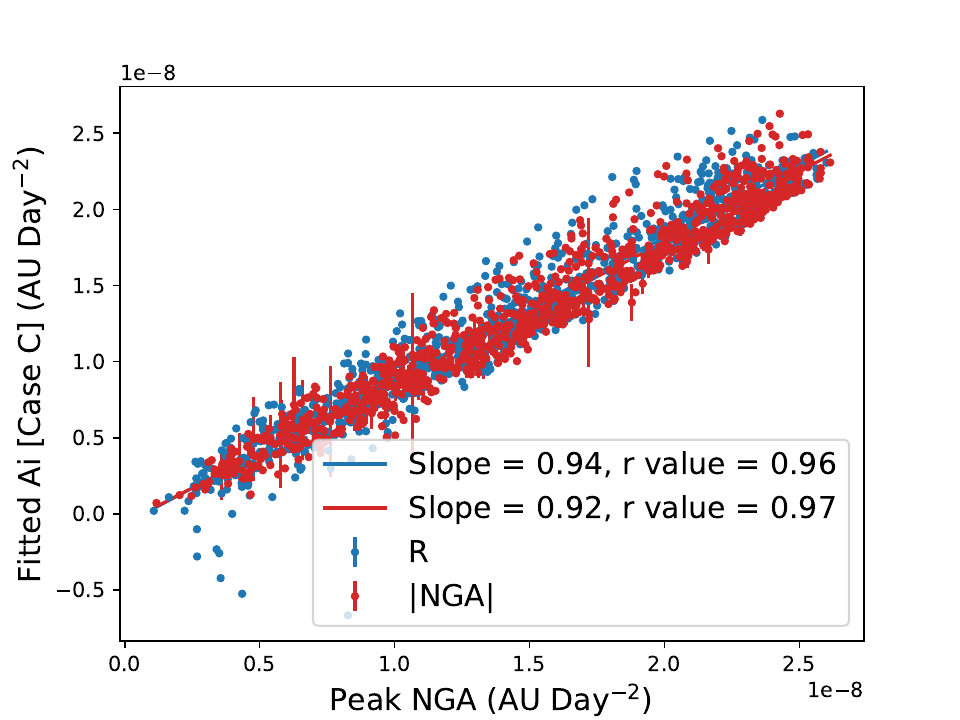}}%
\caption{Fitted acceleration parameters (radial component $A_{1}$ and total magnitude, $A$) as a function of peak thermophysical model accelerations (radial component $\vec{\mathbf{a}}_{R}$ and total magnitude $|\vec{\mathbf{a}}|$) for comet C/2001 Q4. Top: Case B, middle: Case C, pre-perihelion data only, bottom: Case C, full orbital-arc data. Error bars are the $1\sigma$ formal errors from the POD.}
\label{fig:NGAvsA1}
\end{figure}

Figure \ref{fig:NGAvsA1} shows the peak magnitude of each of our Monte Carlo sampled Model II thermophysical acceleration curves versus its Case B and Case C fitted $A_{i}$ values. Here, we show the radial components ($ \vec{\mathbf{a}}_{R}$ and $A_{1}$) and total magnitudes ($|\vec{\mathbf{a}}|$ and $A=\sqrt{A_{1}^{2}+A_{2}^{2}+A_{3}^{2}}$) for comet C/2001 Q4 (the other two comets have nearly identical trends). There is a clear correlation between the peak thermophysical radial acceleration and the peak fitted radial acceleration and, since the radial component dominates the total acceleration, between the magnitudes as well. The correlation exists for both Case B (top) and Case C (middle), but is weaker for Case C due to the instability inherent in asymmetric NGA modelling leading to a number of failed fits and outlying points. There is also a difference in the slope of the correlation: asymmetric Marsden $A_{1}$ terms are on average smaller than symmetric ones. When we consider the full-arc solutions for Case C (bottom), we can see that the outlier points are reduced, and the slope of the correlation now approaches one, i.e.~agreement between the thermophysical peak accelerations and the fitted asymmetric Marsden terms.

These correlations are promising due to the fact that Marsden acceleration magnitude, $A$, is often used alongside measurements of outgassing to determine comet masses via simple conservation of momentum considerations. Fitted $A$ values correlating with true acceleration peaks supports this method, although care must always be taken with uncertainties \citep{Spada}, and the fact that the time of peak acceleration may not necessarily correspond to that of peak outgassing. From the full-arc analysis, we can also see that the best fit to the true underling accelerations is achieved when using an asymmetric Marsden model with the full data arc including pre- and post-perihelion observations. 

We additionally find a strong correlation between the peak NGA magnitude (and hence, the radial component) and the peak water outgassing. Meanwhile, an analysis of the Case A errors showed them to be positively correlated with radial NGA. It is therefore tempting to suggest that the predictability of an incoming comet's orbit could be directly related to its brightness, i.e.~brighter comet $\rightarrow$ larger outgassing $\rightarrow$ larger radial NGA $\rightarrow$ larger trajectory error at intercept. In reality, however, each of these correlations is limited, first by the simplicity of our outgassing/brightness model, and second by the complications induced by the more accurate orbit fitting models in Case B and C.

We now consider the relation between the other Marsden terms ($A_{2}$, $A_{3}$, and $\tau$) and the tangential and normal acceleration components, and time-offset of the peak in total acceleration, $|\vec{\mathbf{a}}|$.

\begin{figure}
\resizebox{\hsize}{!}{\includegraphics[scale=0.5]{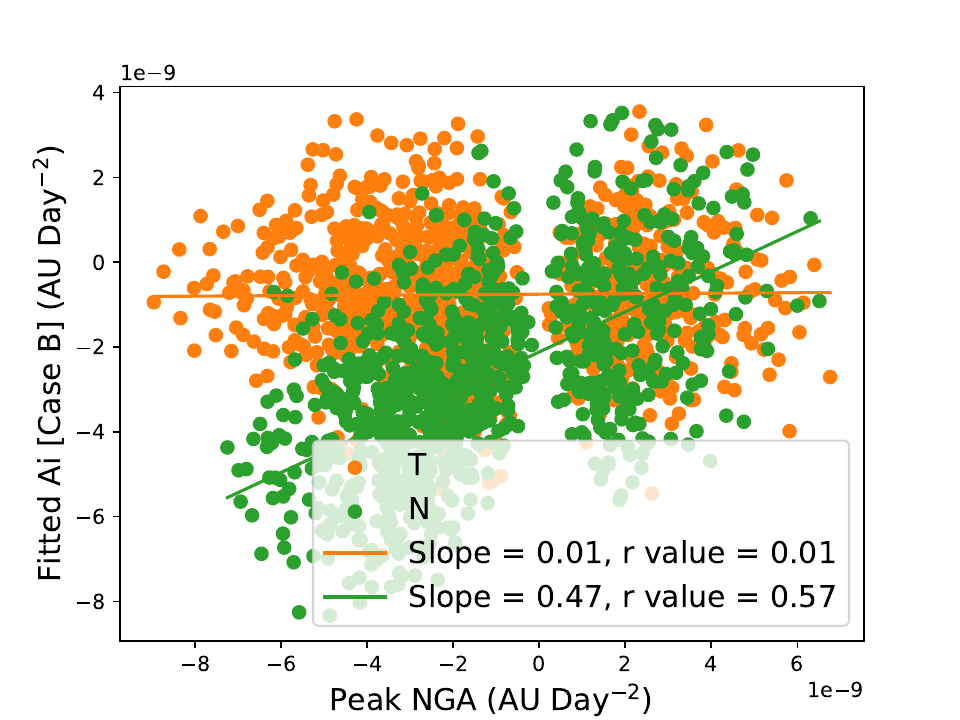}}%
\qquad
{\includegraphics[scale=0.5]{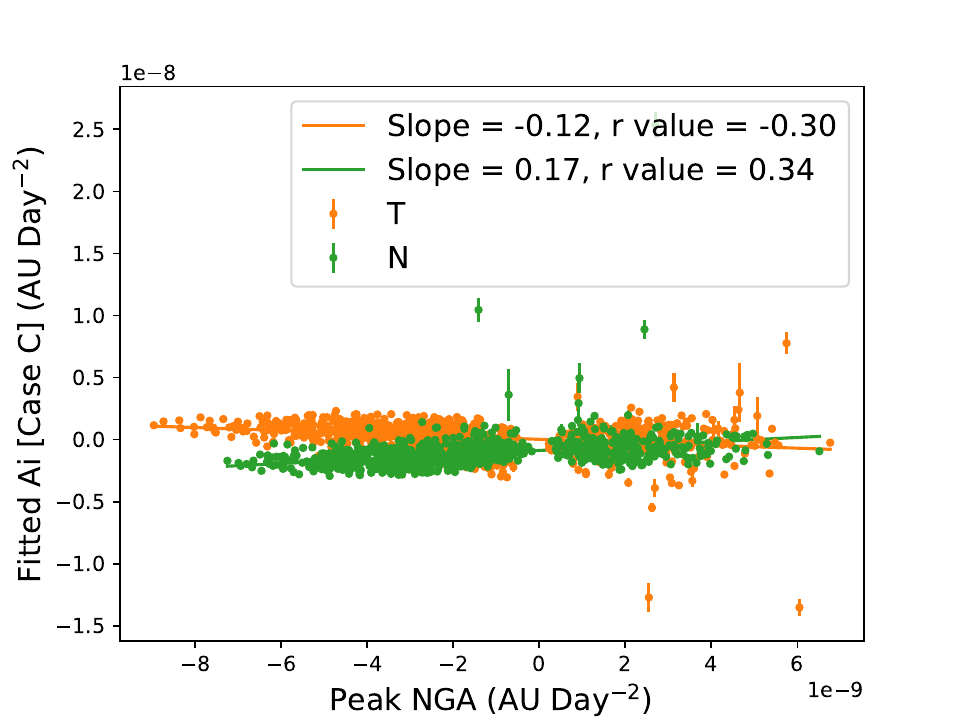}}%
\caption{Fitted acceleration parameters (tangential and normal components $A_{2}$ and $A_{3}$) as a function of peak thermophysical model accelerations (tangential and normal components $\vec{\mathbf{a}}_{T}$ and $\vec{\mathbf{a}}_{N}$) for comet C/2001 Q4. Top: Case B, bottom: Case C.}
\label{fig:NGAvsAi}
\end{figure}

Figure \ref{fig:NGAvsAi} shows correlations between the fitted terms and the physical quantities that can be positive or negative, with varying magnitudes, depending on the acceleration component and NGA model being examined. Similarly unclear results are seen for the other two comets. This poor identification of $A_{2}$ in out fits stands in contrast to typical periodic-comet orbit models, where the tangential component is usually the best-estimated Marsden parameter. This is due to its association with the easily measured change in orbital period \citep{Sekanina1993}, which is particularly clear in the symmetric case. For the hyperbolic comets studied here, the lack of well-defined period, and the extreme sensitivity of osculating parameters, such as the inverse semi-major axis and eccentricity, to differences in non-gravitational fits \citep{Krolikowska2001} blurs this otherwise straightforward correlation. Conversely, our fits appear most sensitive to the radial component $A_{1}$, through its effects on the perihelion argument and time of passage, $\omega$ and $T$. This is also true in Case C, but here the asymmetry allows radial and tangential accelerations to both influence each of these osculating elements, `spreading' the acceleration between them, naturally reducing $A_{1}$ with respect to symmetric case, where it is overestimated (Fig.~\ref{fig:NGAvsA1} and previous paragraph). When analysing the full-arc solutions, we again find weak correlations between the other fitted and underlying acceleration components; showing that, as compared to the improvement for the radial component above, tangential and normal accelerations are always harder to fit for hyperbolic comets.

\label{sec:appendix_correlations}

\begin{figure}
    \centering
    \includegraphics[width=1\linewidth]{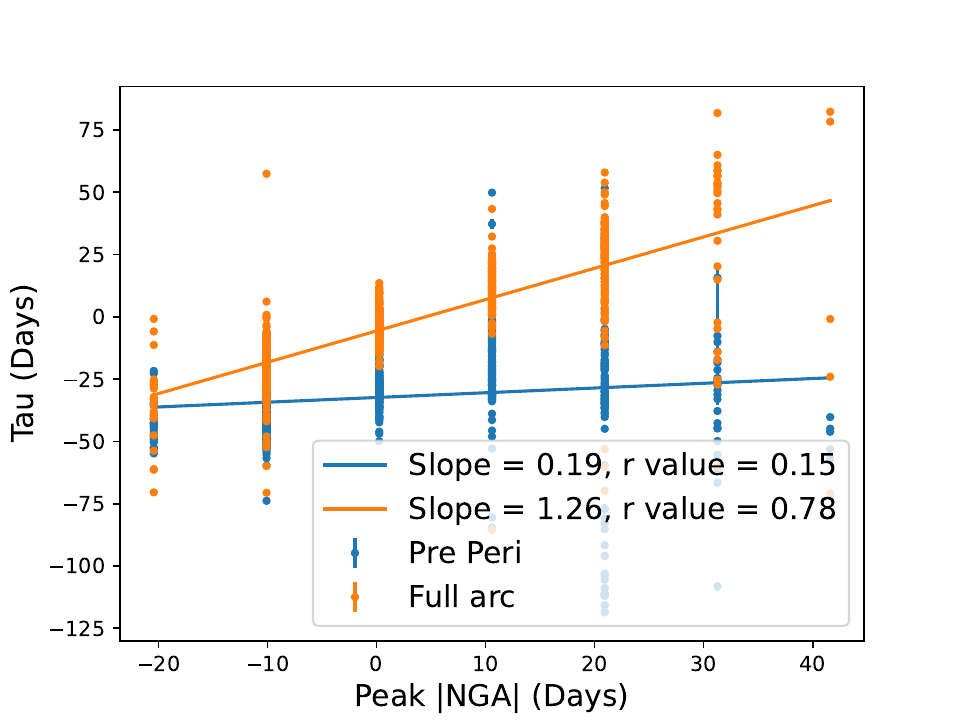}
    \caption{Fitted acceleration parameter (peak NGA time-offset from perihelion, $\tau$) as a function of time of the peak in thermophysical model total acceleration $|\vec{\mathbf{a}}|$ for comet C/2001 Q4. The limiting of peak NGA times to certain values is due to the time-resolution of the thermophysical model.}
    \label{fig:NGAtvstau}
\end{figure}

Figure \ref{fig:NGAtvstau} shows the relation between the fitted and physical time-offsets; these should be correlated according to the original asymmetric model of \citet{yeomans-1989}. Such a correlation is indeed observed for the full orbital-arc solutions (orange points), albeit with significant scatter that can lead to estimated time-offsets on the opposite side of perihelion from the underlying accelerations peaks.

With only pre-perihelion data points, however, there is even more scatter and very little correlation between $\tau$ and the peak thermophysical acceleration (for comet  C/2001 Q4; for the other two comets there is even a weak negative correlation). Fitted time-offsets are nearly always negative, i.e.~before perihelion, even for thermophysical accelerations that peak outbound. That the model is unable to fit the time of an acceleration peak based on data excluding said peak is perhaps not surprising, and likely goes some-way towards explaining the poor accuracy and outlying $A_{i}$ terms of many Case C fits. Indeed, for C/2001 Q4, pre-perihelion fits with post-perihelion $\tau$ values typically had lager formal errors for all parameters, and produced larger errors at perihelion, than those that peaked before perihelion. This suggests that an \textit{a priori\it} limit of $\tau<0$ may help when fitting inbound comet orbits, although the trends were less clear-cut for the other comets, and the median error values change little when applying this cut. Overall then, particular comet orbits must be treated carefully on an individual basis, with careful consideration of their brightening/acceleration trends and the observing arc.

The above analysis shows that Marsden terms, both symmetric and asymmetric, can be interpreted in terms of real physical quantities of the NGA curves of DN LPCs in a statistical sense. Individual comparisons must be done with caution, however, due to the large and often highly-correlated uncertainties (see again \citealp{Spada}). In particular, we find that the total acceleration magnitude and the radial component can be estimated even with only pre-perihelion data, and better estimated when post-perihelion data is included. On the other hand, the time-offset in the peak acceleration can only be reliably estimated with a full orbital arc, and the tangential and normal components remain difficult to estimate, even then.

\subsection{Implications for the Comet Interceptor Mission}
\label{Discussion:implications}

We now return to the question at hand: are we able to estimate the orbits of incoming comets perturbed by non-gravitational forces with sufficient accuracy to predict their positions and velocities at perihelion to within 1000 km and 1 ms$^{-1}$, sixty days before? In short; not with Marsden-style NGA-fits and reasonable quality astrometric data. For both Marsden-style and thermophysical reference trajectories (Models I and II), and a variety of fitting techniques (Cases A, B, and C), typical uncertainties at 60 days are of the order $\sim10,000$ km in position and a few metres per second in velocity. These difficulties stem from a combination of factors such as astrometric accuracy, observation arc, and cadence (not investigated here), orbital geometry, NGAs, and NGA models.

Each of these factors appear to influence the final predicted accuracy, with differences between the three studied comet orbits manifesting themselves differently with the various fitted NGA models. This demonstrate the interconnectedness of the factors, in-that geometric considerations interplay with NGAs and NGA models (if a comet's particular orbit orientation aligns one of the RTN vectors with a particular heliocentric direction, for example) as well as observing constraints (if this then maps onto a particular vector relative to Earth, particularly for the difficult-to-observe out-of-plane components, for example). Increasing the accuracy of observations, and the length of the observation arc clearly helps improve the orbit determinations (compare comet  C/2008 A1 to the other two, for example), but the limitation of the arc to only the pre-perihelion period is not found to be a critical factor.

From the current analysis, it is difficult to ascribe relative proportions to each of the factors. For this, a significant analysis of the correlations within the fitted covariance matrices is required, most likely for an expanded and more uniform range of geometric orbit and NGA orientations. Future work is planned to investigate this. 

A strong factor, however, is clearly the perturbation from NGAs, which typically only becomes measurable in exactly the same time-frame as considered here ($\sim100 - 50$ days before perihelion). There is some slight cause for optimism here as we have taken the time for interception as perihelion itself to simplify the above calculations, whereas in reality Comet Interceptor will target a node crossing \citep{jones-2024}, which could occur some short time before. NGAs may be slightly reduced in their effects at this time, therefore, but the difficulty in predicting their perturbation around perihelion from large heliocentric distance data is clearly demonstrated. A particular comet's acceleration is dependent of many factors (including its orbit, nucleus orientation and shape, and intrinsic activity due to volatile content, thermal properties, regolith depth/active spots, etc.~), which cannot be estimated ahead of time and are very difficult to predict into the future. NGAs then must be studied in a statistical sense, hence the Monte Carlo approach here, and gradually refined with more accurate models. Once such model, not studied here, is the rotating-jet model \citep{Chesley2005}, which has been successfully applied to improve fly-by predictions for incoming LPCs (\citealp{Farnocchia2016}: in this case, for the planet Mars). Rotating-jet models also raise the possibility of extracting additional information about a comet's spin orientation, at the cost of adding significant complexity to the analysis, and therefore will be the subject of future study. Another possibility that deserves future study could be an arc-wise parametrisation of the Marsden model.

Clearly, some sort of mitigation is necessary if Comet Interceptor is to be successful. The ESOC flight dynamics team, who provided the original accuracy requirement, are taking into account the $\sim10,000$ km-level uncertainties at sixty days described above when planning the approach to encounter phase that begins here (ESOC, personal communications). During this phase the spacecraft's own NAVCAM (and potentially the COmetCAmera science instrument) will be used to track the target comet which, alongside radio-tracking of the spacecraft, will provide enormously-enhanced astrometric precision for trajectory predictions. See, for example, improvements to the final Giotto-Halley flyby precision \citep{Curdt1988} from optical navigation and tracking of the preceding Vega missions \citep{Muench1986}, as well as recent analysis of interstellar object interception by \citet{Mages2023}.

It is possible that early observations of the target comet with NAVCAM and/or CoCa, from well-before the sixty day close-approach phase, could significantly improve the orbit determination accuracy at the start of this phase. Another potential source of additional astrometric information will be observations from off-Earth. The Hubble and James Webb Space Telescopes, as well as solar system missions at Mars (Mars Express and the ExoMars Trace Gas Orbiter) and in deep space (JUICE on the way to Jupiter), have recently been used to significantly improve orbit estimations of interstellar object 3I/ATLAS \citep{Eubanks2025}, for example, and it is likely similar attempts would be made with a potential Comet Interceptor target. Both of these techniques could be simulated using appropriate synthetic observations and the method developed above, and therefore will be the subject of future work.

In summary then, some improvements may be possible with sources of additional astrometric information and higher fidelity NGA models, but this remains subject to future modelling. For now, Comet Interceptor's successful rendez-vous with an LPC is not challenged by the analysis, but it has been shown that NGAs pose a significant effect that must be taken into account during the planning and navigation processes. One final area of relevance for Comet Interceptor is the large error seen in the first orbit determination after object discovery (stemming purely from the difficulties of gravitational orbit determination at this point) and its swift decline with additional observations. This was not investigated in detail here as we focus on encounter navigation after initial target selection. Future work, however, could use the above methods to quantify the impact of high initial uncertainty on the timing of the target selection process \citep{snodgrass-2025}.

\section{Conclusions}
\label{sec:conclusions}

We have established a general framework for evaluating the predictability of incoming long-period comet trajectories under non-gravitational perturbations. Although developed in support of the Comet Interceptor mission, the methodology is applicable to any problem requiring quantitative assessment of comet ephemeris uncertainty, including future survey operations, mission target selection, and studies of cometary dynamics. We first create a reference trajectory, including NGAs via either parametrised Marsden models, or a Monte Carlo sampled thermophysical model. We then generate synthetic observations of the reference trajectory, before fitting orbits to increasing data arcs. Finally we sample the associated covariance matrices and propagate the uncertain orbits forward to the chosen intercept point in order to compare with the original reference trajectory. We applied this method to three long-period comets previously selected as virtual targets for the Comet Interceptor mission. A number of conclusions can be drawn from the results:

\begin{itemize}
    \item Precise orbit determination for incoming LPCs perturbed by NGAs is unlikely to be able to achieve the accuracy specified in design documents for the Comet Interceptor mission ($1000$ km and 1 m s$^{-1}$ in position and velocity, sixty days before encounter) when modelling the NGAs with Marsden-style parameters. Instead, we recommend planning for $10,000$ km and a few times 1 m s$^{-1}$.
    \item This uncertainty comes from a combination of the limitations of the astrometric observations, and our ability to predict how NGAs evolve, and varies on a case-by-case basis; depending on a particular comet's orbital geometry, Earth-centric range, observability, as well as its intrinsic activity and rotation (e.g.~two equally well-observed comets might have different predictability depending on their NGAs).
    \item Asymmetric Marsden models (the highest-fidelity type of NGA model tested here) provide the best results for realistic NGAs, with significant improvements over pure-gravity orbit determinations, and minor improvements over symmetric Marsden models.
    \item Both types of Marsden model struggle to accurately predict trajectories, however, when there is limited information available on the NGAs (when they are just 'turning on' at a few AU), and may provide worse results here than pure-gravity solutions.
    \item Despite this, fitting only half an orbital arc, i.e.~up until perihelion, as compared to both pre- and post-perihelion, is found to be able to achieve reasonable predictive accuracy at perihelion, depending on the specific orbit.
    \item Fitted-Marsden terms, such as the peak magnitude of the acceleration and its radial component, are found to be positively correlated with their corresponding underlying physical quantities. This is broadly supportive of the extraction of physical quantities from NGA fits (e.g.~mass and radius; \citealp{sosa-2011, Eubanks2025}, etc.~), even for incoming comets (previous point); however, acceleration time-offsets can only be estimated with full-arc data. Caution must also be taken with errors, given the large scatter and correlated uncertainties \citep{Spada}.
\end{itemize}

The last point (that serious consideration must be paid to both the overall magnitude of the covariance matrix attached to an orbit determination, and also correlations within it) also applies to the intercept prediction accuracy. As shown for the three example LPCs examined here, orbit fitting is fiddly and complicated, and must be approached on a case-by-case basis with tight control on both \textit{a priori} and observational uncertainties, and a healthy degree of scepticism for the derived quantities and their formal errors.

In future work, we plan to expand this method to a wider sample of orbital geometries, with more systematic control over the input variables, in order to gauge the relative importance of observations, orbits, and NGAs to the final prediction accuracies. We also plan to incorporate off-Earth observations (from space telescopes, solar system missions, and Comet Interceptor itself), in order to test their power to enhance the orbit determination, as well as higher-fidelity NGA models, such as with a rotating jet.

\section*{Acknowledgements}

N.A. and P.G. acknowledge financial support from project PID2021-126365NB-C21 (MCI/AEI/FEDER, UE) and from the Severo Ochoa grant CEX2021-001131-S funded by MCI/AEI/10.13039/501100011033. P.L. was supported in part by the \emph{ISSI} Visiting Scientist program, while P.d-B's internship was supported by ERASMUS+. The authors also wish to thank Colin Snodgrass, Abbie Donaldson, and the other members of the Comet Interceptor working groups for useful discussions and input that helped develop this work, as well as the reviewer, Declan Mages, for comments and suggestions that improved the text.

\section*{Data Availability}
The data underlying this article will be shared on reasonable request to the corresponding author.




\bibliographystyle{mnras}
\bibliography{paper_references} 


\appendix 
\section{Velocity Uncertainties}
\label{sec:appendix_velocities}

\begin{figure}
    \centering
    \includegraphics[width=1\linewidth]{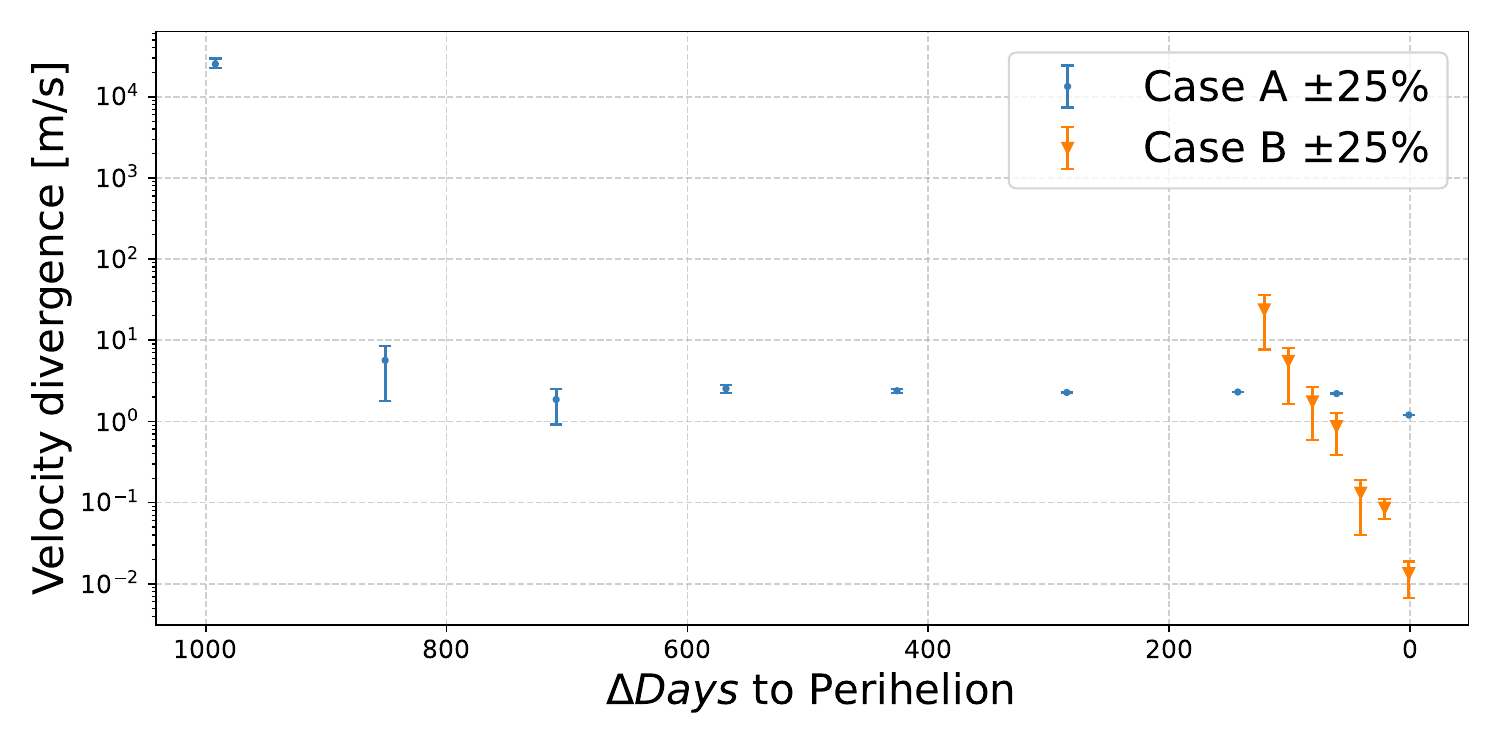}
    \caption{Perihelion velocity distribution of fitted comet C/2001 Q4 orbits compared to a Marsden reference orbit (Model I). The x-axis describes when the last observation took place, the y-axis describes the distribution at perihelion, and the points show medians with error bars given by the interquartile range. Blue dots are Case A and orange triangles, Case B.}
    \label{fig:Marsden_2001Q4v}
\end{figure}
\begin{figure}
    \centering
    \includegraphics[width=1\linewidth]{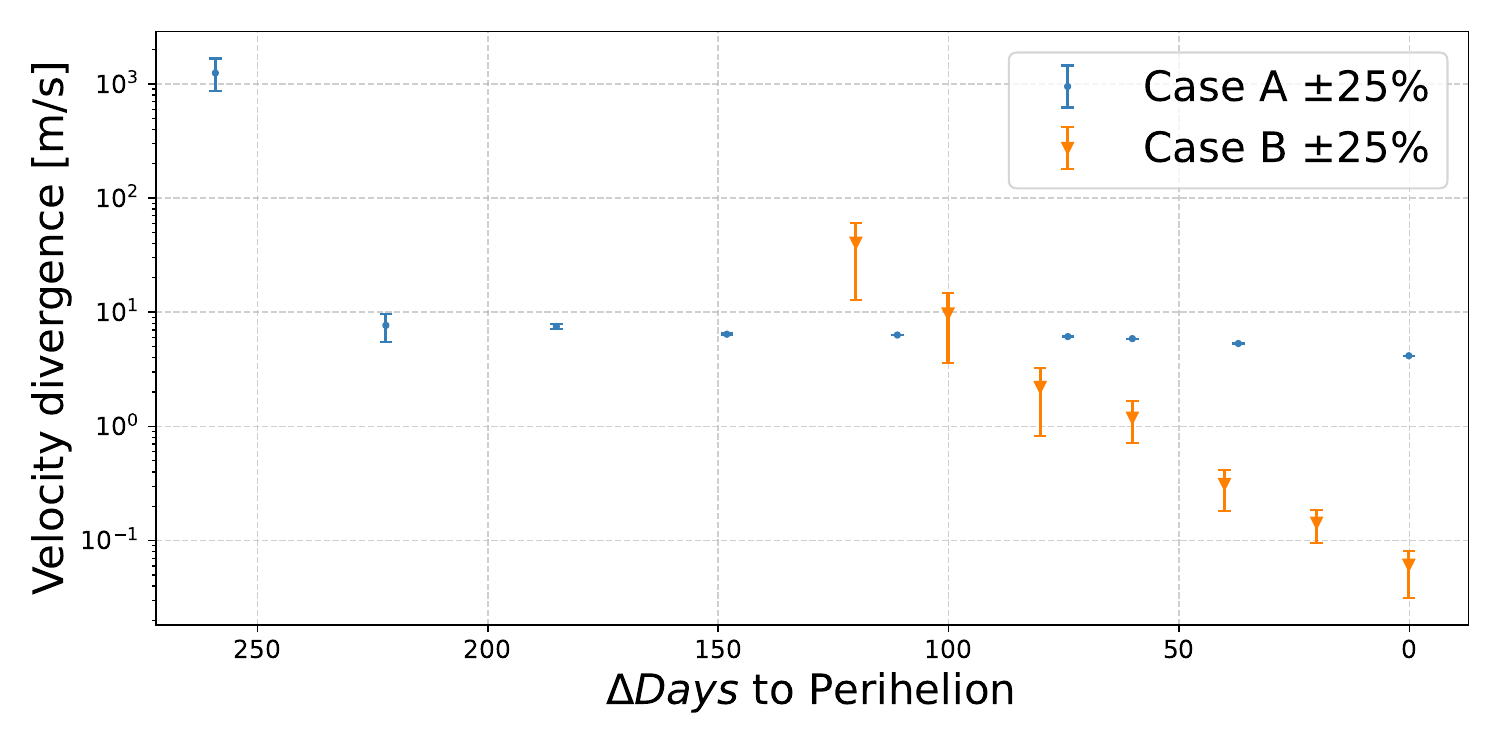}
    \caption{Perihelion velocity distribution of fitted comet C/2008 A1 orbits compared to a Marsden reference orbit (Model I). The points and axes are as above in Fig.~\ref{fig:Marsden_2001Q4v}.}
    \label{fig:Marsden_2008Av1}
\end{figure}
\begin{figure}
    \centering
    \includegraphics[width=1\linewidth]{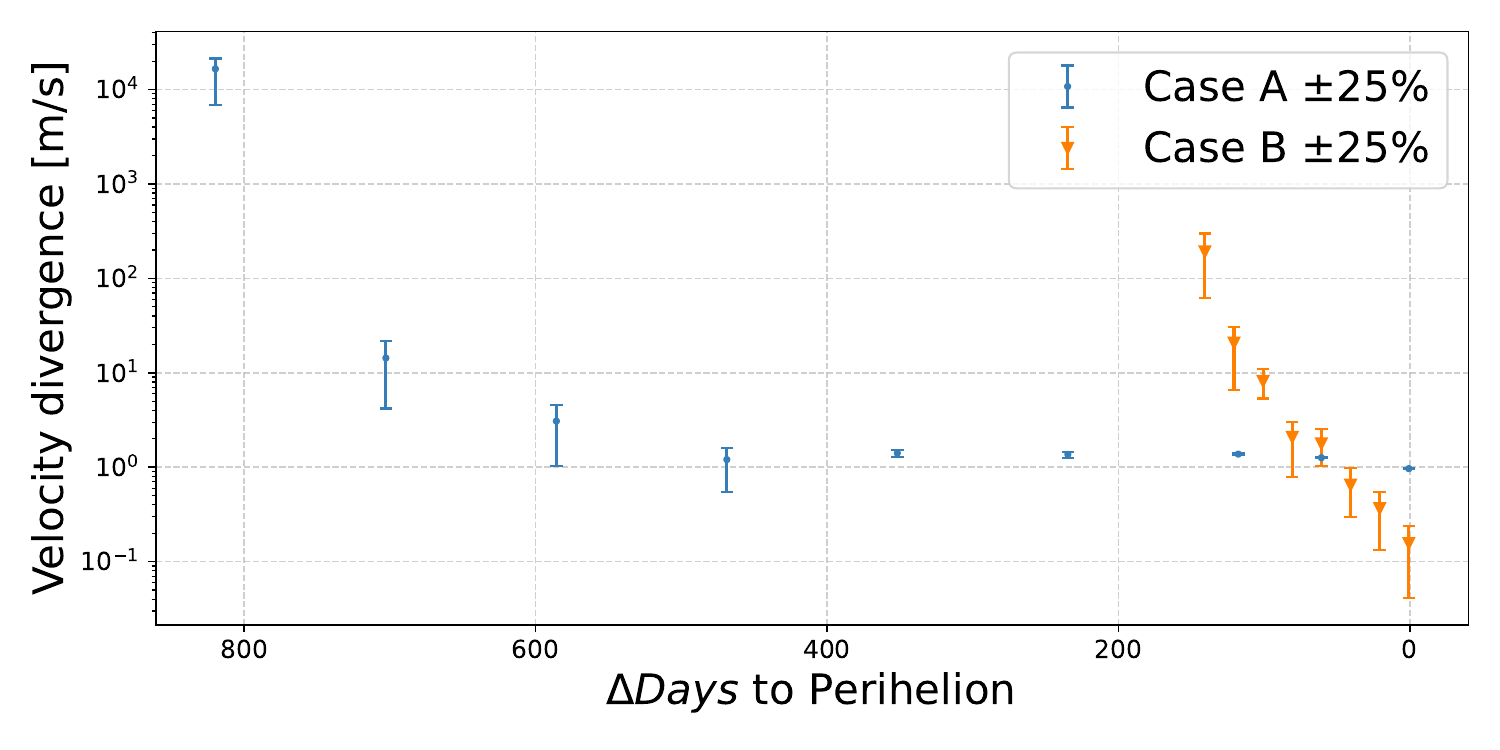}
    \caption{Perihelion velocity distribution of fitted comet C/2013 US10 orbits compared to a Marsden reference orbit (Model I). The points and axes are as above in Fig.~\ref{fig:Marsden_2001Q4v}.}
    \label{fig:Marsden_2013US10v}
\end{figure}

\begin{figure}
    \centering
    \includegraphics[width=1\linewidth]{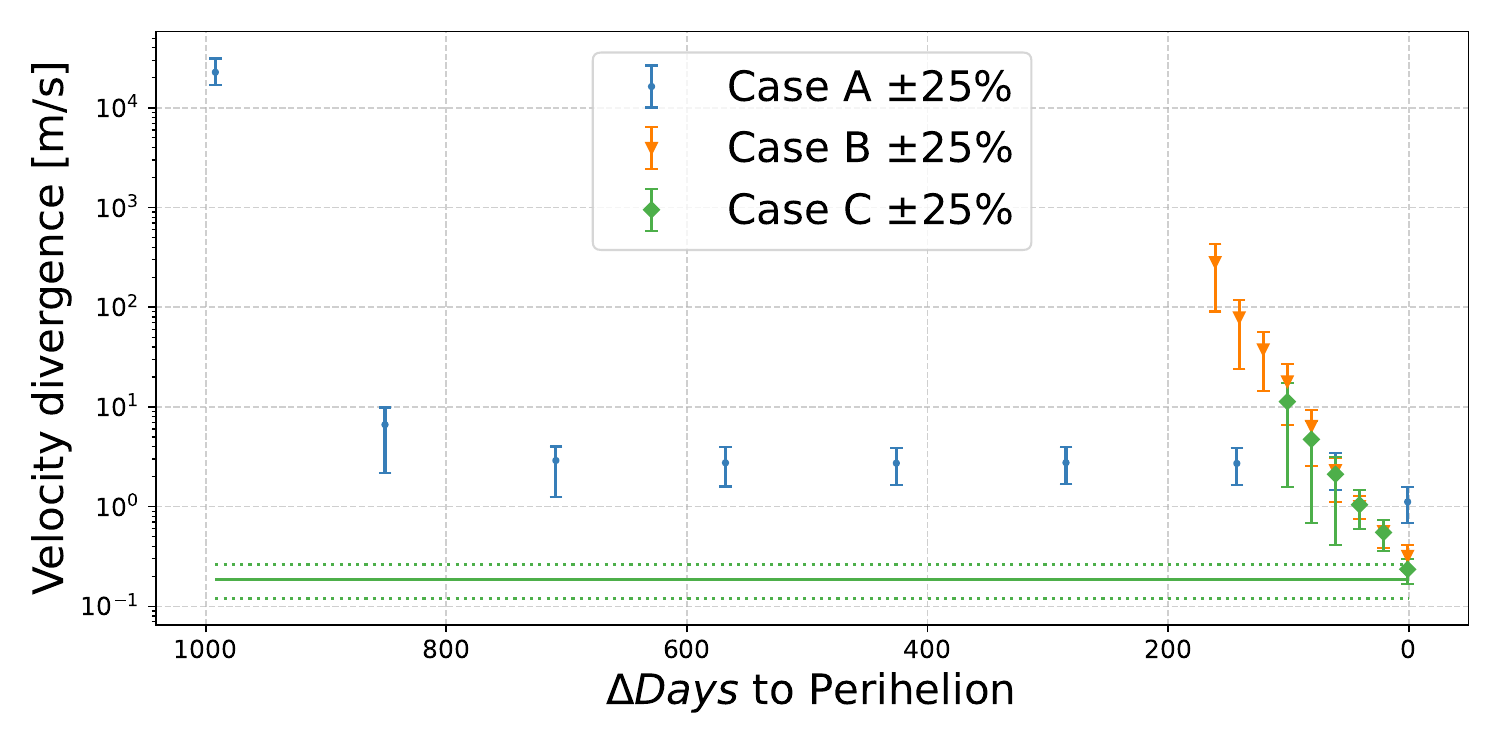}
    \caption{Perihelion velocity distribution of fitted comet C/2001 Q4 orbits compared to thermophysically-perturbed reference orbits (Model II). The x-axis describes when the last observation took place, the y-axis describes the distribution at perihelion, and the points show medians over the full suite of Monte Carlo samples with error bars given by the interquartile range. Blue dots are Case A, orange triangles, Case B, and green diamonds, Case C. The horizontal solid and dotted lines denote the full-orbit Case C median and interquartile range respectively.}
    \label{fig:Thermal_2001Q4v}
\end{figure}
\begin{figure}
    \centering
    \includegraphics[width=1\linewidth]{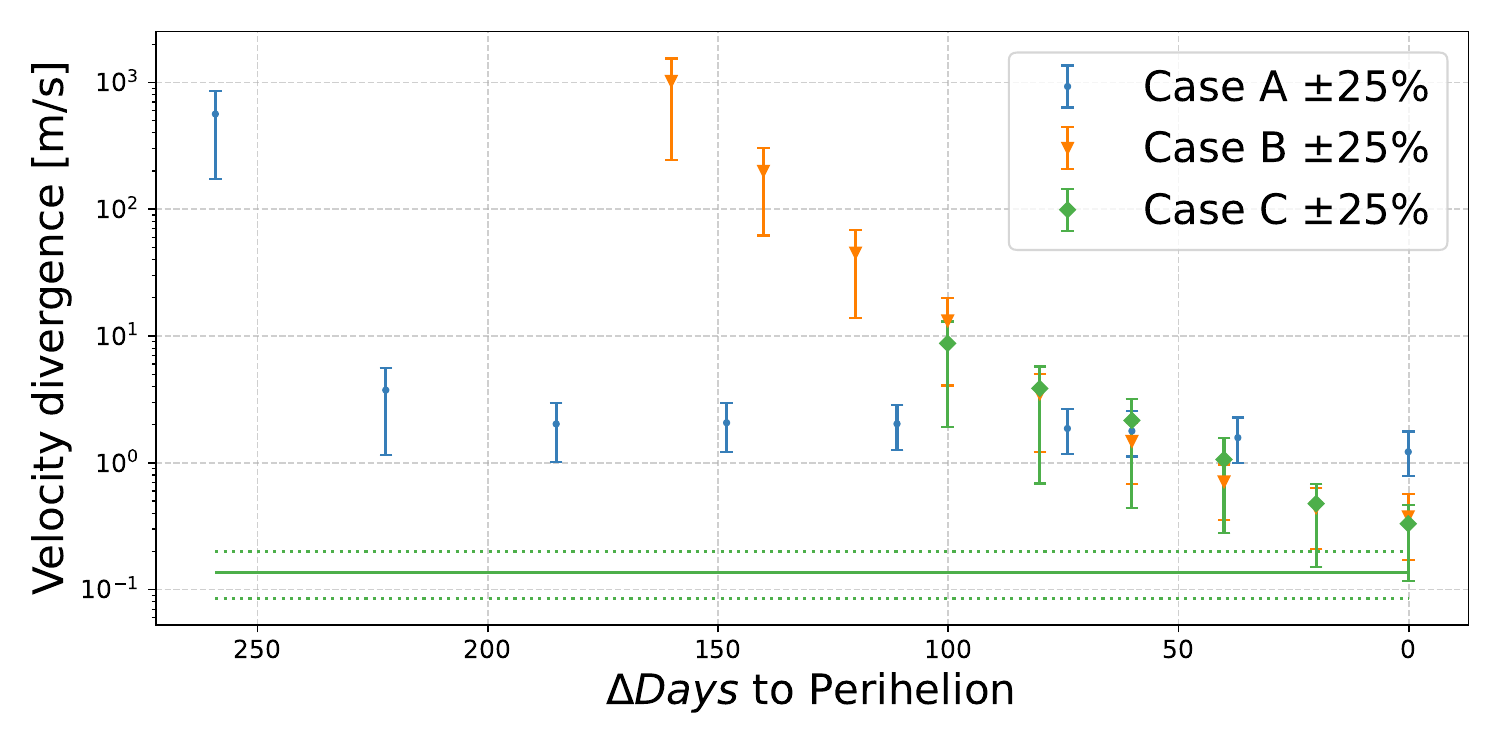}
    \caption{Perihelion velocity distribution of fitted comet C/2008 A1 orbits compared to thermophysically-perturbed reference orbits (Model II). The points and axes are as above in Fig.~\ref{fig:Thermal_2001Q4v}.}
    \label{fig:Thermal_2008A1v}
\end{figure}
\begin{figure}
    \centering
    \includegraphics[width=1\linewidth]{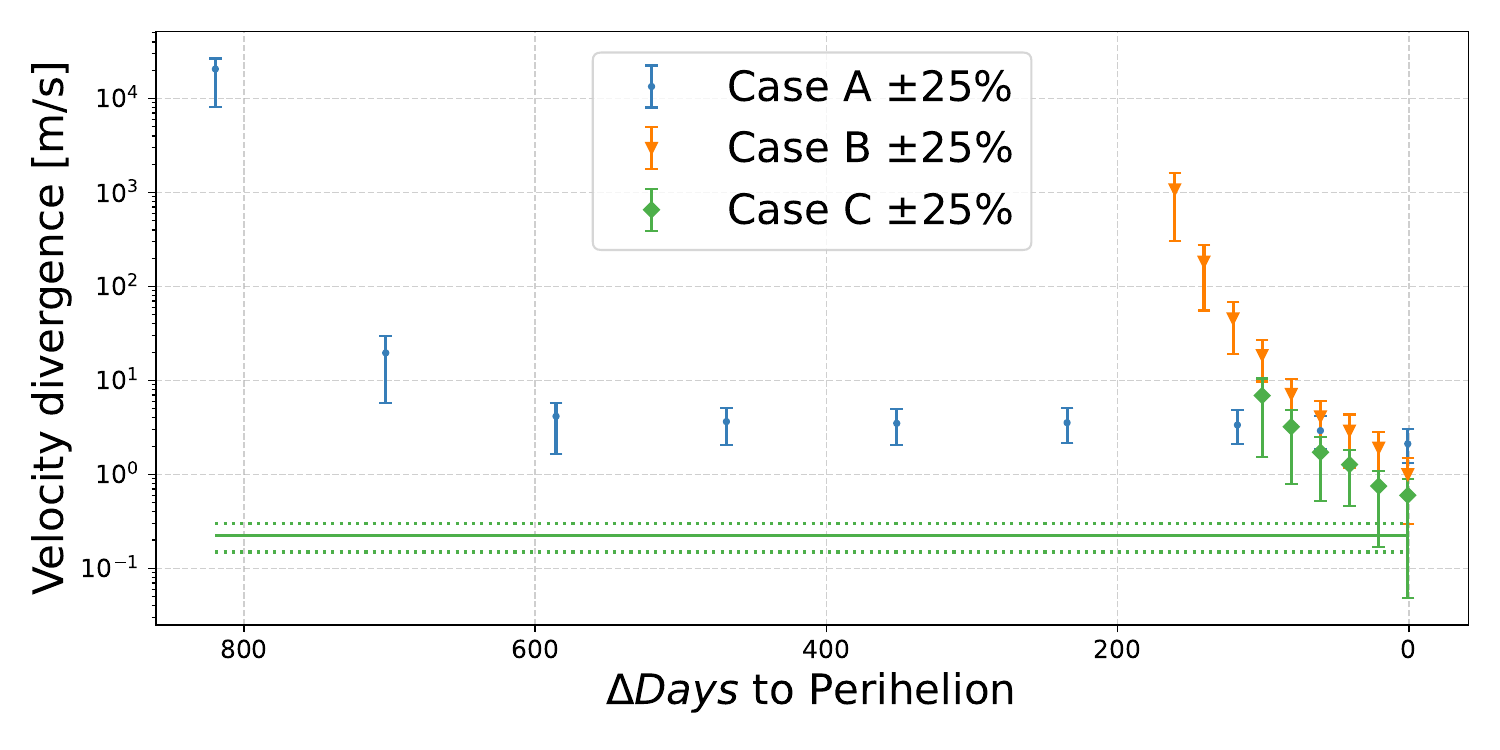}
    \caption{Perihelion velocity distribution of fitted comet C/2013 US10 orbits compared to thermophysically-perturbed reference orbits (Model II). The points and axes are as above in Fig.~\ref{fig:Thermal_2001Q4v}.}   
    \label{fig:Thermal_2013US10v}
\end{figure}

\section{Effects of increased astrometric noise}
\label{sec:appendix_noise}

\begin{figure}
    \centering
    \includegraphics[width=1\linewidth]{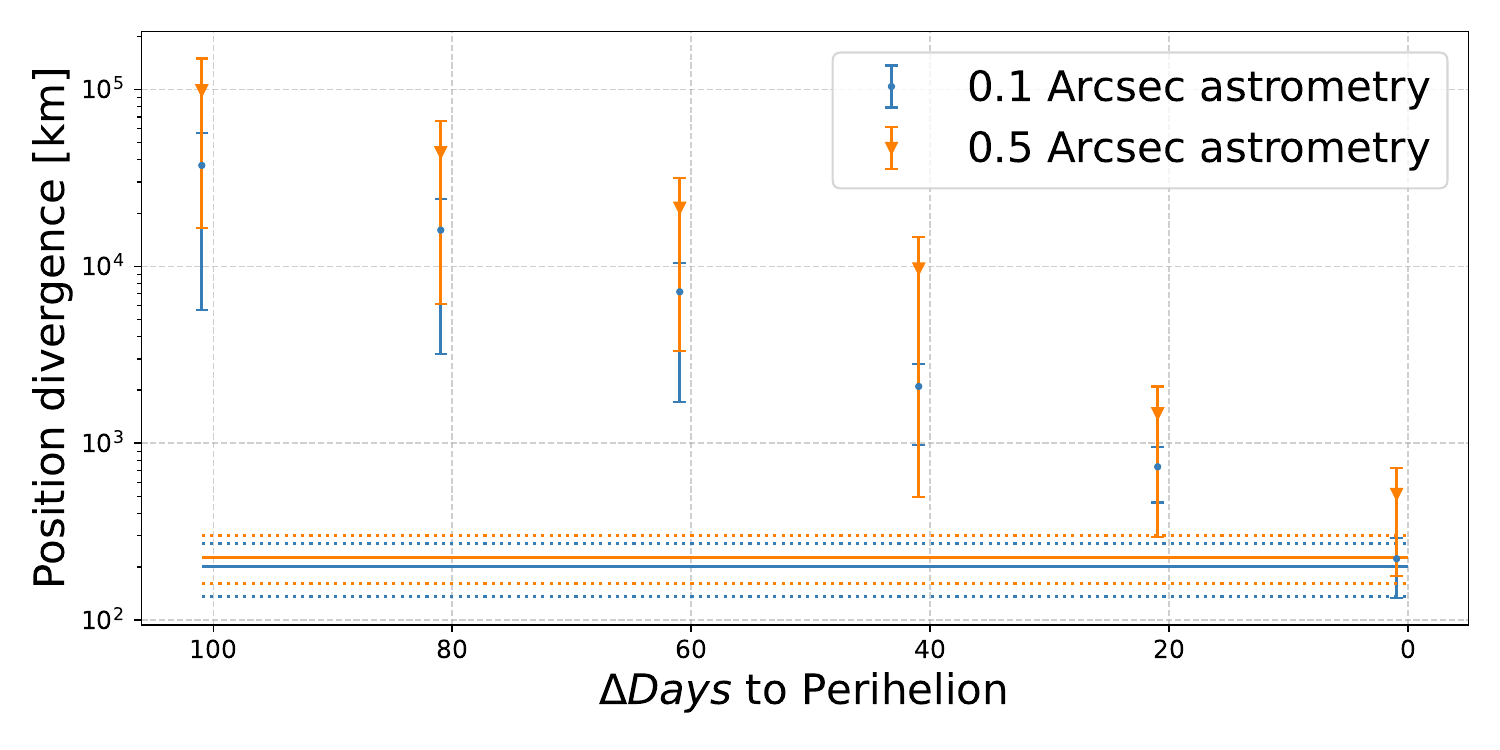}
    \caption{Perihelion position distributions, $\Delta$, of fitted comet C/2001 Q4 orbits with increased astrometric noise compared to thermophysically-perturbed reference orbits (Model II). The x-axis describes when the last observation took place, the y-axis describes the distribution at perihelion, and the points show medians over the full suite of Monte Carlo samples with error bars given by the interquartile range. Blue dots are Case C with the nominal astrometric noise, while orange triangles have it increased by five times. The horizontal solid and dotted lines denote the full-orbit Case C median and interquartile range respectively.}
    \label{fig:Thermal_2001Q4_05}
\end{figure}
\begin{figure}
    \centering
    \includegraphics[width=1\linewidth]{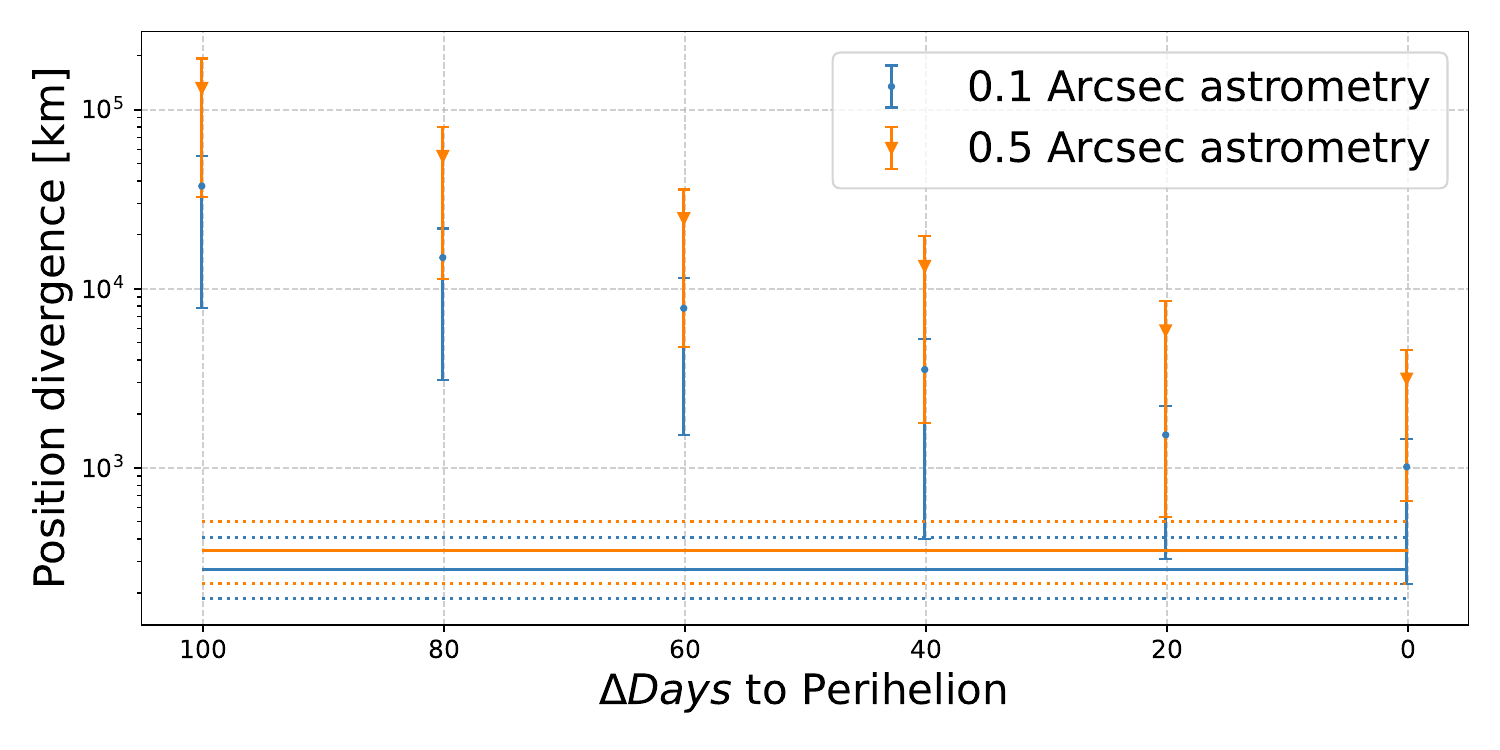}
    \caption{Perihelion position distributions, $\Delta$, of fitted comet C/2008 A1 orbits with increased astrometric noise compared to thermophysically-perturbed reference orbits (Model II). The points and axes are as above in Fig.~\ref{fig:Thermal_2001Q4_05}.}
    \label{fig:Thermal_2008A1_05}
\end{figure}
\begin{figure}
    \centering
    \includegraphics[width=1\linewidth]{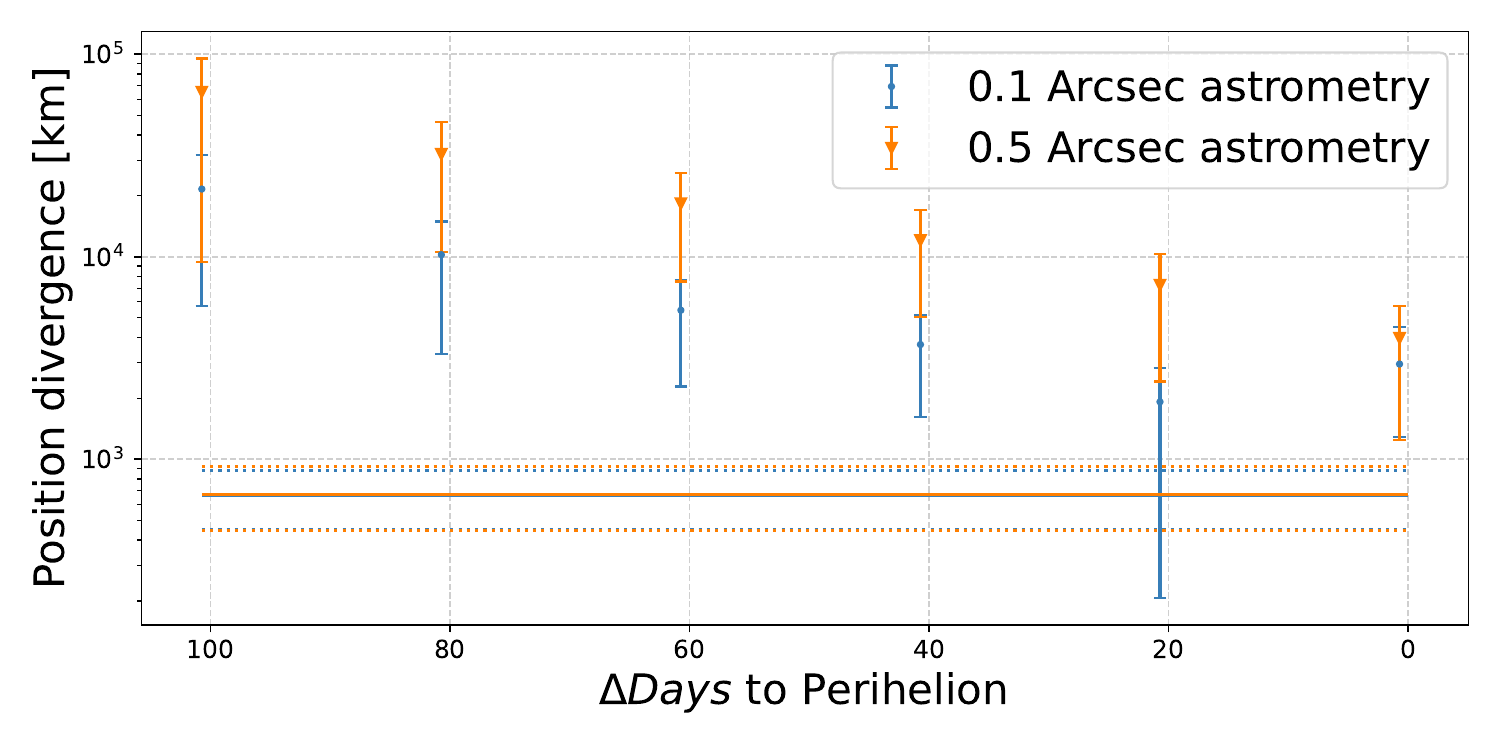}
    \caption{Perihelion position distribution of fitted comet C/2013 US10 orbits with increased astrometric noise compared to thermophysically-perturbed reference orbits (Model II). The points and axes are as above in Fig.~\ref{fig:Thermal_2001Q4_05}.}   
    \label{fig:Thermal_2013US10_05}
\end{figure}

\bsp	
\label{lastpage}
\end{document}